\documentclass[aps,pra,twocolumn,superscriptaddress]{revtex4-1}

\usepackage{bm}
\usepackage[retainorgcmds]{IEEEtrantools}
\usepackage{graphicx}
\usepackage{mathrsfs}
\usepackage{amsmath}
\usepackage{amssymb}
\usepackage{color}
\usepackage{amsfonts}
\usepackage{nicefrac,txfonts}
\newcommand{\up}{\uparrow}
\newcommand{\down}{\downarrow}
\newcommand{\Tr}{{\rm Tr}}

\begin{document}

\title{Excitation-transport in open quantum networks: The effects of configurations}
\author{James Cormican}
\affiliation{Centre for Theoretical Atomic, Molecular and Optical Physics, School of Mathematics and Physics, Queen's University Belfast, Belfast BT7 1NN, United Kingdom}
\author{Lorenzo Stella}
\affiliation{Atomistic Simulation Centre, School of Mathematics and Physics, Queen's University Belfast, Belfast BT7 1NN, United Kingdom}
\affiliation{School of Chemistry and Chemical Engineering, Queen's University, Belfast BT7 1NN, United Kingdom}
\author{Mauro Paternostro}
\affiliation{Centre for Theoretical Atomic, Molecular and Optical Physics, School of Mathematics and Physics, Queen's University Belfast, Belfast BT7 1NN, United Kingdom}
\affiliation{Laboratoire Kastler Brossel, ENS-PSL Research University, 24 rue Lhomond, F-75005 Paris, France}
\date{\today}

\begin{abstract}
We study the environment-assisted enhancement of the excitation-transport efficiency across a network of interacting quantum particles or sites. 
Our study reveals a crucial influence of the network configuration --- and especially its degree of connectivity --- on the amount of environment-supported enhancement. In particular, we find a significant interplay of direct and indirect connections between excitation-sending and receiving sites. 
On the other hand, the non-Markovianity induced by memory-bearing, finite-size environments does not seem to provide a critical resource for the enhanced excitation-transport mechanism. %The relevance of our findings in the study of quantum-enhanced excitation-transport in biological systems or artificial nanostructures is finally discussed [is that true?]}  

\end{abstract}
\maketitle{}

\section{Introduction}	

It has long been recognized that, under suitable conditions, noise could be not just a source of detrimental effects for the dynamics of a system, but a mechanism for potential advantages. A significant, well-known example in classical physics is provided by stochastic resonance~\cite{Gammaitoni, Haenggi}, the phenomenon according to which, the signal-to-noise ratio of a given non-linear process can be enhanced by the addition of moderate-intensity white noise, owing to the occurrence of resonances.  

The benefits of noise in quantum dynamics, however, are not only less evident but also more subtle.  
Large environmental fluctuations typically act as sources of decoherence for a quantum system, destroying quantum coherences and leading to classical stochastic processes. Dissipation, on the other hand, depletes the performance of the {excitation-transport mechanisms}. On one hand, these phenomena have inspired a technological race to realization of conditions for ultra-high isolation of quantum systems from the corresponding environment. The results of such a race are the remarkable progresses achieved so far in the quantum control of the dynamics of a large number of quantum systems, from trapped ions to cold-atom and superconducting quantum devices, which are making the goal of a realizing a quantum processor foreseeable in a close future. On the other hand, the unavoidable system-environment interaction has inspired research aimed at understanding if and how environmental effects can be turned into an advantage and lead to the creation of quantum coherence in a given system. Remarkably, examples have been given of such a possibility~\cite{Plenio1,Hartmann,Sink1,Huelga}, in particular when addressing the dynamics of quantum networks that could model the working mechanisms of biological systems.

Very significant steps have been taken towards the understanding of the features that provide an environment-assisted enhancement of the performance of {excitation-transport} in photosynhetic quantum networks~\cite{Sink1,Chin,HuelgaPlenio,OlayaCastro}. The role of quantum interference operated by modest amounts of environmental dephasing or dissipation appears to be crucial, in this respect~\cite{Checinska}. It is also important to remark the interest generated by system-environment interactions in light-harvesting systems for the production and control of quantum correlations~\cite{Caruso,Sarovar,Bradler,Cifuentes}. 

Yet, much remains to be understood of the phenomenology of environment-assisted enhancement of the excitation-transport, from the influences of environmental memory effects to the significance of quantum coherence. Among the questions that are currently open, is the dependence of the performance of excitation-transport on the {network configuration}. This is a particularly relevant point when addressing quantum effects in light-harvesting systems, {given the} large degree of connectivity of their underlying networks~\cite{HuelgaPlenio}, and for the engineering of {excitation-transport} in nanostructures~\cite{OlayaCastro}, which might benefit of specifically arranged network configurations. Needless to say, the effect of network configuration on the efficiency of excitation-transport processes represents a question that can be addressed beyond the boundaries of the specific contexts mentioned above, namely quantum biology and the engineering of artificial nanostructures. {Indeed, this problem would benefit} of an abstract approach assessing the performance of general networks of connected sites and undergoing non-equilibrium open-system dynamics. 

This is precisely the viewpoint that we take in this paper, where we study the interplay between the configuration of a network of interacting {particles} and the occurrence of environment-assisted quantum transport (ENAQT). We first model a process of excitation-transport in a multi-site graph open to both local and collective environments. We then unveil an intriguing trade-off between the enhancement of the excitation-transport efficiency in the presence of modest amounts of dephasing noise, and the existence of multiple interfering pathways for the excitation-transport from a designated sending site to a receiving one. {We show that existence of a direct link between the sending and receiving sites is crucial for the achievement of enhanced ENAQT.} 

We also assess the effects that non-Markovianity induced by network-environment interaction might have on the enhanced ENAQT effect that we discuss. We thus compare the excitation-transport performance when the environments with which the network is in contact are memoryless --- thus enforcing a standard Markovian dynamics on the network --- and when they are able to induce a non-Markovian evolution. Our results show that non-Markovianity does not appear to be a resource {for ENAQT}, a conclusion that we verify through an extensive numerical assessment of the network dynamics. {Our numerical findings reinforce} the idea that {graph-connectivity} is a key factor in establishing enhanced environment-induced effects. This provides valuable information on the way, for instance, a nanostructure should be engineered to operate advantageously in the presence of an environment, which can be useful towards the construction of quantum-enhanced nanoscale devices and processes.

%Open quantum systems experience noise due to external environment. 
%Such noise comes in many different forms, from dephasing to pure dissipation, which in geenral spoil the quantum features of a given system. However, hese are potentially powerful as while removing noise is quite often a difficult task, controlling a certain boundary of noise is often relatively easy. With this in mind, we have been considering the improvement of excitation-transport across a system through the action of a dephasing channel. This excitation-transport phenomenon is called Environmental-Assisted Quantum Transport (ENAQT). We consider the potency and accessibility of this effect under various environmental conditions.
%\ls{More about your findings and non-markovianity (see conclusions)}

The remainder of this paper is structured as follows: In Sec.~\ref{system} we introduce the formal description of the system that we consider, introduce the models for environmental mechanisms, and the figure of merit for the quantification of the effects that we aim at exploring. Sec.~\ref{analysis} presents our analysis of the dynamics, highlighting the situations that underpin {the sink excitation probability (SEP)} and the occurrence of ENAQT. 
Our study is quite extensive, and covers the effects of non-Markovianity as well as the influence that quantum coherences have on the phenomenology that we illustrate. Finally, Sec.~\ref{conclusions} summarizes our findings and draws their implications for the investigation of noise-affected quantum dynamics in quantum networks. 

\section{The system and its modelization}
\label{system}
\begin{figure}[b!]
\includegraphics[width=0.9\linewidth]{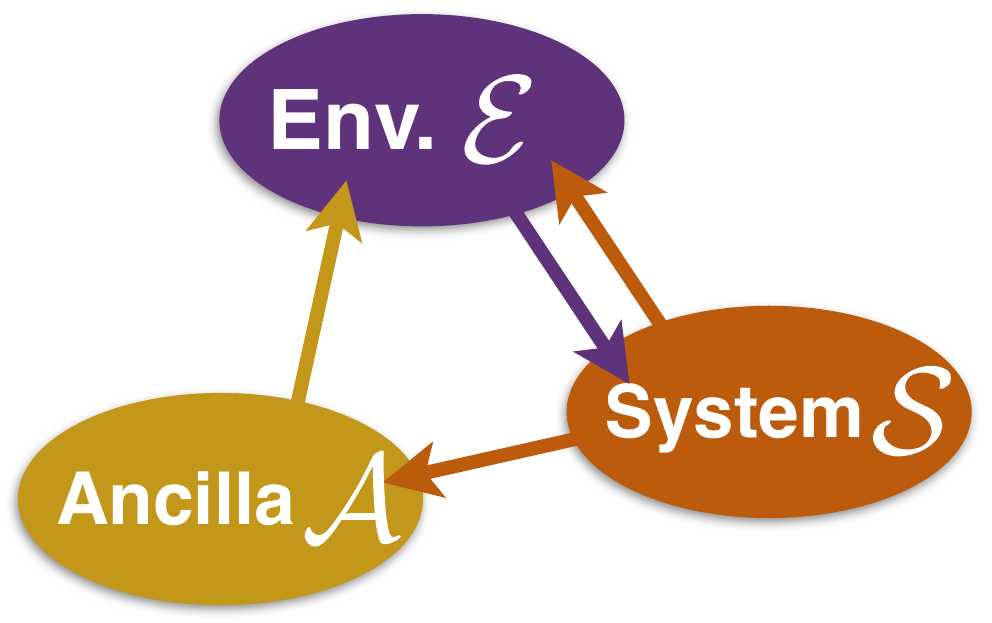}
\caption{Sketch of the configuration of interactions among system ${\cal S}$, ancilla ${\cal A}$, and environment ${\cal E}$. Both system and ancilla are, in general, multipartite. In particular, we allow for a set of intra-system couplings, whose form is determined by the adjacency matrix ${\bm \tau}$, and ${\cal S}-{\cal A}$ interactions, as specified by matrix ${\bm q}$. The effects of ${\cal E}$ on both the ancilla and the system is treated phenomenologically by assuming memoryless (Markovian), independent environments. Not included in the sketch is the sink used to trap the excitations that, having been transported across ${\cal S}$, reach a given site of the system. }
\end{figure}
We consider a system of interacting qubits (hereby dubbed {\it the system} and labeled as ${\cal S}$) coupled to individual and independent environments and thus undergoing open-system dynamics. Besides the intra-system coupling and the interaction with the surrounding environment, we also consider an (in general multi-qubit) ancillary system (dubbed {\it the ancilla} and labeled as ${\cal A}$) that, in principle, undergoes open dynamics as well. 

As our primary goal is to quantify the excitation-transport efficiency from a sender to a receiving qubit across the system, we incoherently couple the latter to a {\it sink} particle. The coupling is devised in a way that the sink would thus absorb excitations reaching the receiving particle without being able to feed such excitations back to the system. Models with artificial sinks could be envisioned as systems that transfer energy to a zero-temperature bath. Similar sinking mechanisms have been employed in the modelization of excitation-transport across photosynthetic and light-harvesting complexes~\cite{HuelgaPlenio} 

The Hamiltonian for a general configuration of the model at hand thus reads (we assume units such that $\hbar=1$ and measure energy in units of the Bohr frequency $\omega$ of the system's and ancilla's particles)
\begin{equation}
\label{hamiltonian}
\begin{aligned}
H(J,Q)&=\sum_{i\in\{{\cal S}\}}\sigma _z^i+\sum_{\alpha\in\{{\cal A}\}}\sigma _z^k+\sum _{i,j\in\{{\cal S}\}}\tau_{ij}(J){\bm \sigma}^i\cdot{\bm\sigma}^j\\%\left ( \sigma_x^i\sigma_x^\alpha +\sigma_y^i\sigma_y^j+\sigma_z^i\sigma_z^j \right ) \\
&+\sum _{i\in\{{\cal S}\}}\sum _{\alpha\in\{{\cal A}\}}q_{ik}(Q){\bm \sigma}^i\cdot{\bm\sigma}^\alpha.%\left ( \sigma_x^i\sigma_x^k +\sigma_y^i\sigma_y^k+\sigma_z^i\sigma_z^k \right )
\end{aligned}
\end{equation}
Here roman (Greek) indices run over the elements of system ${\cal S}$ (the ancilla ${\cal A}$) and ${\bm \sigma}^{k}=(\sigma^k_x,\sigma^k_y,\sigma^k_z)$ is the vector of Pauli matrices of particle $k=i,\alpha$. We have introduced the adjacency matrix for the intra-system couplings ${\bm \tau}(J)$ and that for the  coherent ${\cal S}$-${\cal A}$ interaction ${\bm q}(Q)$. The strength of the respective coupling is set by the dimensionless rates $J$ and $Q$. Both ${\bm \tau}(J)$ and ${\bm q}(Q)$ are used here to control the details of the configuration of interactions within the model. Specifically, we have the following adjacency matrix entries % These topological control matrices have the following form;
\begin{equation}
    \tau_{ij}(J)=
\begin{cases}
    J& \text{when qubits $i,j\in{\cal S}$ are coupled},\\
    0              & \text{otherwise}
\end{cases}
\end{equation}
and
\begin{equation}
    q_{i\alpha}(Q)=
\begin{cases}
    Q& \text{when qubit $i\in{\cal S}$ is coupled to} \\
       &\text{qubit $\alpha\in{\cal A}$},\\
    0              & \text{otherwise}.
\end{cases}
\end{equation}
%The strength of the respective coupling is set by the dimensionless rates $J$ and $Q$. %Note that this Hamiltonian describes the global system (primary system and ancilary system).The Lindblad Master Equation of our model has the following form;

We now include the effects of the environment (labeled as ${\cal E}$), which we assume to consist of individual memoryless mechanism affecting each particle of the system (and possible the ancilla) independently. The {joint} dynamics of ${\cal S}$ and ${\cal A}$ {is assumed} to be Markovian and {modeled by} phenomenological Lindblad-like superoperators. In order to {describe the general case}, we {consider the environment} ${\cal E}$ {as including} both dephasing and dissipative effects on the system. {These effects are modeled by the superoperators}
\begin{equation}
\begin{aligned}
{\cal L}_{\text{deph}}(\rho)&=D\sum _{i\in{\cal S}}\left ( \sigma_i^z \rho \sigma_i^z-\rho \right),\\
{\cal L}_{\text{damp}}(\rho)&=\gamma _\up \sum _{i\in{\cal S}}N_i\left ( \sigma_i^+ \rho \sigma_i^--\frac{1}{2}\left \{\sigma_i^-\sigma_i^+, \rho \right \} \right )\\
&+\gamma _\down\sum _{i\in{\cal S}}(N_i+1)\left ( \sigma_i^- \rho \sigma_i^+-\frac{1}{2}\left \{\sigma_i^+\sigma_i^-, \rho \right \} \right ).
\end{aligned}
\end{equation}
Here $\rho$ is the joint ${\cal S}$-${\cal A}$ {density matrix},  $D$ is the {dephasing rate} (assumed to be uniform across the system), $\gamma_{\down}$ ($\gamma_\up$) is the rate of incoherent loss (incoherent pump) of excitations into (from) the environment attached to particle $i\in{\cal S}$.
{The environment is assumed to be in thermal equilibrium} at temperature $T_i$ and with an average number of excitations $N_i=(e^{\beta_i}-1)^{-1}$, where ${\beta_i=1/K_BT_i}$ is inverse temperature and $K_B$ is the Boltzmann constant.

In our model, the ancilla has two {configurations}: individual and communal. In the individual one, the constituents of the ancillary system do not interact with each other and are individually coupled to one part of ${\cal S}$. In the communal configuration, the ancillary system interacts with the entire ${\cal S}$ at once. In both cases we allow for individual non-zero temperature damping effects on ${\cal A}$ of the form % $E_A$ describes the incoherent coupling between the ancillary system and the primary system and has the following form;
%$E_{AI}$ describes a coupling between the Markovian environment and the ancillary system.
\begin{equation}
\begin{aligned}
{\cal L}_{{\cal A}}(\rho)&=\gamma^A_\up N_A\sum_{\alpha\in{\cal A}}\left ( \sigma_\alpha^+ \rho \sigma_\alpha^--\frac{1}{2}\left \{\sigma_\alpha^-\sigma_\alpha^+, \rho \right \} \right )\\
&+\gamma^A _\down(N_A+1)\sum_{\alpha\in{\cal A}}\left ( \sigma_\alpha^- \rho \sigma_\alpha^+-\frac{1}{2}\left \{\sigma_\alpha^+\sigma_\alpha^-, \rho \right \} \right ),
\end{aligned}
\end{equation}
where $N_A$ is the average number of excitations in the  bath attached to ${\cal A}$ and $\gamma^{A}_{\up,\down}$ are the corresponding rates of incoherent loss and pump.

Finally, we introduce the sink mechanism. As mentioned at the beginning of this Section, this can be understood as damping of excitations into a zero-temperature environment, which is unable to feed them back into the system. {A minimal model for such a sink is given by} a two-level system incoherently coupled to the $n^{\text{th}}$ site of the system with decoherence rate $\gamma_S$, as~\cite{Chin}
\begin{equation}
{\cal L}_S(\rho)=\gamma_S\left(\sigma_S^+ \sigma_n^- \rho \sigma_n^+ \sigma_S^--\frac{1}{2}\left \{ \sigma_n^+ \sigma_S^- \sigma_S^+ \sigma_n^- ,\rho  \right \}\right).
\end{equation}
With this at hand, {the non-unitary evolution of} the ${\cal S}$-${\cal A}$  {density matrix is obtained by solving the equation} 
\begin{equation}
\label{masterequation}
\partial_t\rho{=}{-}i\left [ H(J,Q),\rho  \right ]{+}{\cal L}_{\text{deph}}(\rho){+}{\cal L}_S(\rho){+}{\cal L}_{\text{damp}}(\rho)+{\cal L}_{{\cal A}}(\rho),
\end{equation}
{This} dynamical equation provides the core information on the excitation-transport problem here under scrutiny, whose performance is characterized quantitatively by considering {SEP}, i.e. the probability that an excitation seeded in one of the particles of ${\cal S}$ reaches the site to which the sink is attached. Formally, SEP is defined as
\begin{equation}
\text{SEP}={}_{S}\langle {\up} | \Tr_{{\cal SA}}(\rho) |{\up}\rangle_{S}
\end{equation}
where $\Tr_{\cal SA}$ stands for the partial trace over the degrees of freedom of ${\cal S}$ and ${\cal A}$, which leaves us with the reduced state of the sink only.  SEP is then calculated by performing the  projection of such reduced state onto the excited state $|{\up}\rangle_S$ of the sink. In what follows, for zero-temperature environments, we will consider the situations when ENAQT occurs at the steady-state of the dynamics, and calculate the corresponding value of SEP, which we will then compare to the corresponding value achieved in the absence of dephasing noise. We dub the difference between SEP in these two configurations as SEP improvement (or SEPI) and define it as
\begin{equation}
\text{SEPI(D)}=\text{SEP(D)}-\text{SEP(0)},
\end{equation}
SEPI thus quantifies the degree of improvement in the sink-excitation probability resulting from ENAQT, an effect which we observe through dephasing.
   %We consider the steady state when observing ENAQT effects as that is the point where we see maximum effect as well as it being the easiest point to standardise when making comparisons. 
For environments at non-zero temperature, on the other hand, we would not consider the steady state of the dynamics, as the latter would not exhibit any ENAQT. We will thus resort to a study of the dynamical state.

%We study the features of SEP against the relevant parameters of the system and for various choices of the adjacency matrix ${\bm \tau}(J)$. %In particular, we address the occurrence of ENAQT and the possibility to observe an environment-induced improvement \ls{"improvement" (or enhancement, if you prefer) is a bit vague --- e.g., what is the difference between a negligible improvement and a significant improvement? Here we should refer to the introduction where previous or expected improvements are mentioned (I suppose)} of SEP, which we dub as SEPI, depending on the dynamical conditions of the system. SEPI quantifies the degree of improvement in the sink-excitation probability resulting from ENAQT, an effect which we observe through dephasing. We thus define SEPI as
%\begin{equation}
%\text{SEPI(D)}=\text{SEP(D)}-\text{SEP(0)},
%\end{equation}
%which depends explicitly on the strength of the dephasing mechanisms in the dynamics of the network. 

%\begin{figure}
%{\bf (a)}\\
%\includegraphics[width=0.9\linewidth]{ANCCOM}\\
%{\bf (b)}\\
%\includegraphics[width=0.75\linewidth]{ANCIN}
%\caption{ We consider two regime types, communal (left) and individual (right). We acheive different structures through manipulating the size of the ancillary system and components of the ancillary topological matrix $q$. }
%\end{figure}

Before turning our attention to the characterization of the excitation-transport process, let us discuss an important point. As we aim at assessing the effects that the ancilla has on the performance of excitation-transport, in what follows we shall compare the results obtained through the coherent ${\cal S}$-${\cal A}$ coupling in Eq.~\eqref{hamiltonian} to what is observed by {replacing the coherent ${\cal S}-{\cal A}$ interaction in Eq.~\eqref{hamiltonian} with the incoherent interaction}
\begin{equation}
\label{incoherentSA}
\begin{aligned}
{\cal L}_{{\cal SA}}(\rho)&{=}\sum _{i\in{\cal S},\alpha\in{\cal A}}\left[\Gamma _\up\left ( \sigma_i^+\sigma_\alpha^- \rho \sigma_\alpha^+\sigma_i^-{-}\frac{1}{2}\left \{\sigma_\alpha^+\sigma_i^-\sigma_i^+\sigma_\alpha^-, \rho \right \} \right ) \right.\\
&{+}\left.\Gamma _\down\left ( \sigma_i^-\sigma_\alpha^+\rho \sigma_\alpha^-\sigma_i^+{-}\frac{1}{2}\left \{\sigma_\alpha^-\sigma_i^+\sigma_i^-\sigma_\alpha^+, \rho \right \} \right )\right],
\end{aligned}
\end{equation}
where $\Gamma_\down$ ($\Gamma_\up$) is the rate at which excitations are transferred from (to) the $i^{\text{th}}$ element of the system to (from) the $\alpha^{\text{th}}$ party of the ancilla. Therefore, the dynamics guided by Eq.~\eqref{masterequation} will be contrasted to the one resulting from the master equation
\begin{equation}
\label{masterequation2}
\begin{aligned}
\partial_t\rho&{=}{-}i\left [ H(J,0),\rho  \right ]{+}{\cal L}_{\text{deph}}(\rho){+}{\cal L}_S(\rho)\\
&{+}{\cal L}_{\text{damp}}(\rho)+{\cal L}_{{\cal A}}(\rho)+{\cal L}_{{\cal SA}}(\rho).
\end{aligned}
\end{equation}

%where the non-unitary evolution terms are Lindblad operator terms. These terms have the following form;
%\begin{equation}
%\begin{split}
%D'(\rho)+S'(\rho)=D\sum _i\left ( \sigma_i^z \rho \sigma_i^z-\rho \right )+S\sigma_S^+ \sigma_n^- \rho \sigma_n^+ \sigma_S^- \\
%-\frac{S}{2}\left \{ \sigma_n^+ \sigma_S^- \sigma_S^+ \sigma_n^- ,\rho  \right \} 
%\end{split}
%\end{equation}
%$D$ is the dephasing coupling strength and $S$ is the sink coupling strength. 

\section{Analysis of the dynamics, phenomenology of SEP and occurrence of ENAQT}
\label{analysis}

%\textcolor{blue}{{\bf (MAURO: to be fixed)} For this model, zero temperature environments are described by regimes where $N_i=0$. Non-zero temperature environments are regimes where at least one $N_i>0$. We considered two archetypes of non-zero temperature regimes, Directional and Full. In the 'directional' scheme $N_1>0$ and all other $N_i=0$. This causes the influx of excitations from the environment to be directed through the system. In the 'full' scheme, all $N_i>0$. In additional, we observed that due to the asymmetric coupling of the sink to the system and the sink's otherwise isolated nature, SEP$=1$ for all warm environments. Thus, rather than examining the SEP directly, we consider the time taken for the SEP$=1$ to occur. When the ancillary system is added to the system, a number of potential environment types can occur. These regimes give rise to totally cold environments, partially warm (either through the primary system or through the ancillary system) or completely warm (through both primary and ancillary systems).}

\begin{figure}
\includegraphics[width=\columnwidth]{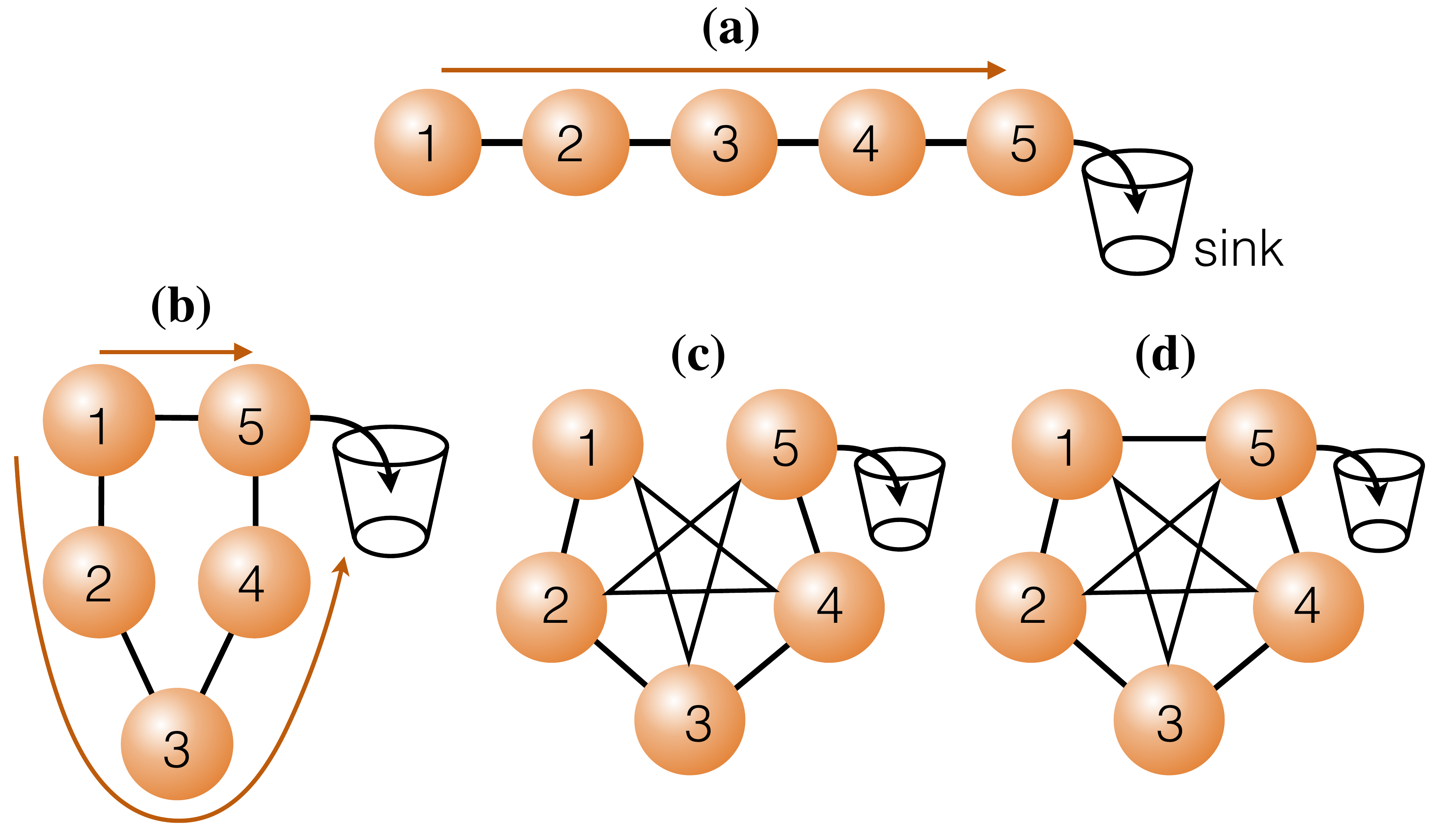}
\caption{Panel {\bf (a)} shows an example of the `linear' configuration: as noted with the arrow, there is one path to the sink from the initial qubit. Panel {\bf (b)} shows an instance of loop configuration with two possible paths to the sink. Panels {\bf (c)} and {\bf (d)} offer a complex variety of possible pathways. We refer to configurations analogous to the one in panel {\bf (c)} as `non-criticalÕ as we observe that ENAQT requires this pathway for the effect to take place. Lastly, we refer to panel {\bf (d)} as `maximally connected' as it contains the maximum number of interactions and thus the maximum number of possible pathways.}
\label{Config}
\end{figure}

\begin{figure*}[ht!]
{\bf (a)}\hskip5cm{\bf (b)}\hskip5cm{\bf (c)}\\
\includegraphics[width=0.67\columnwidth]{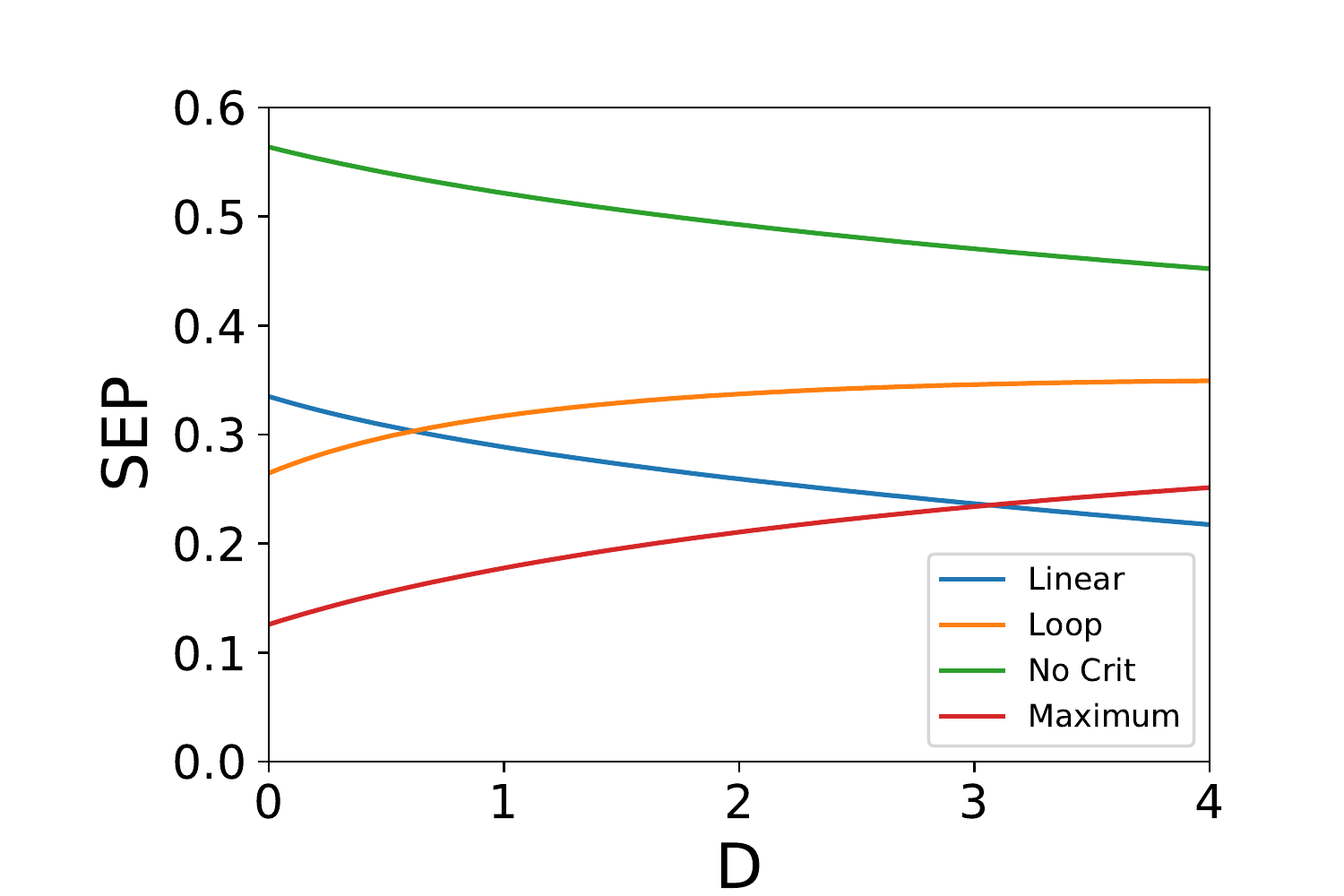}\includegraphics[width=0.67\columnwidth]{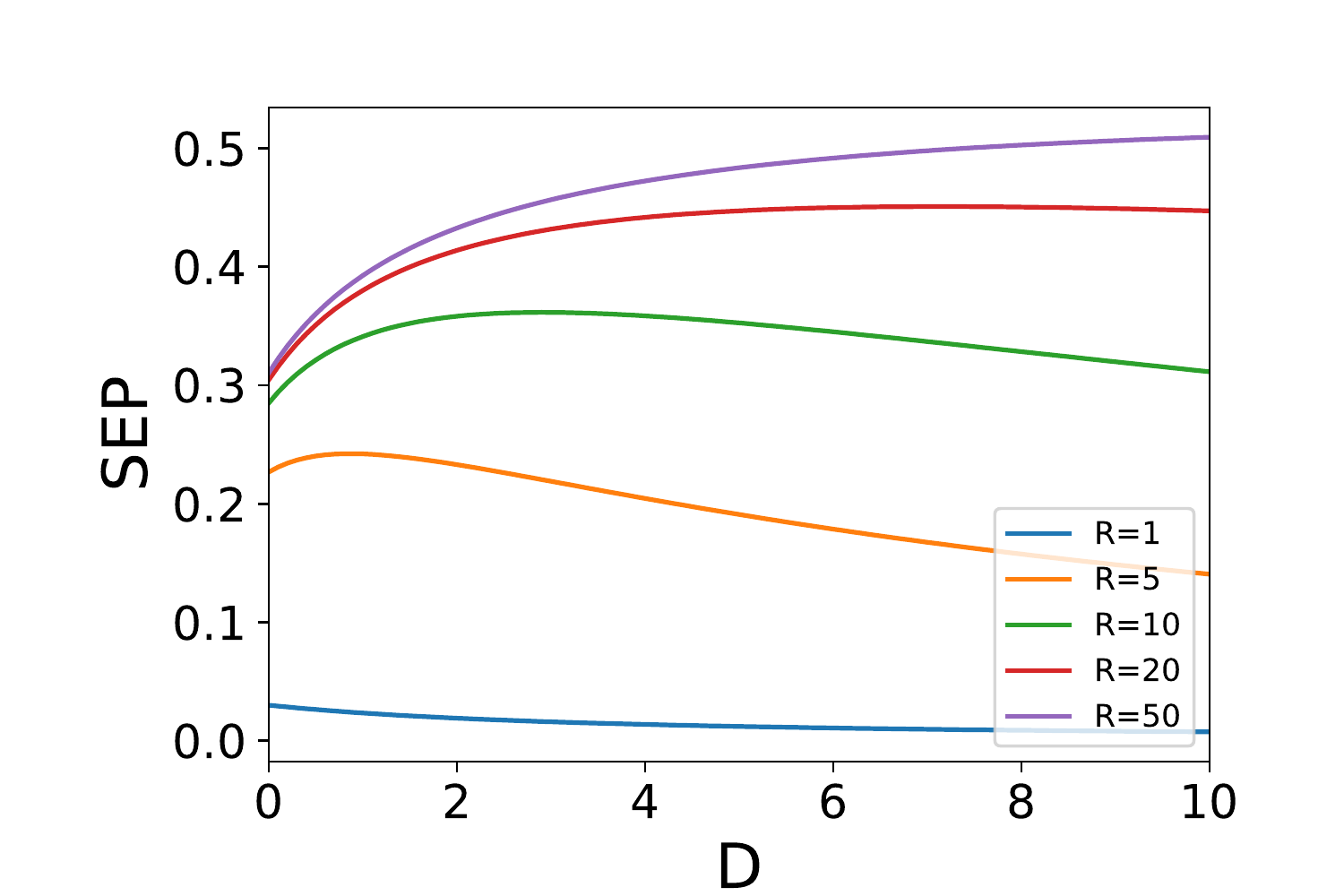}\includegraphics[width=0.63\columnwidth]{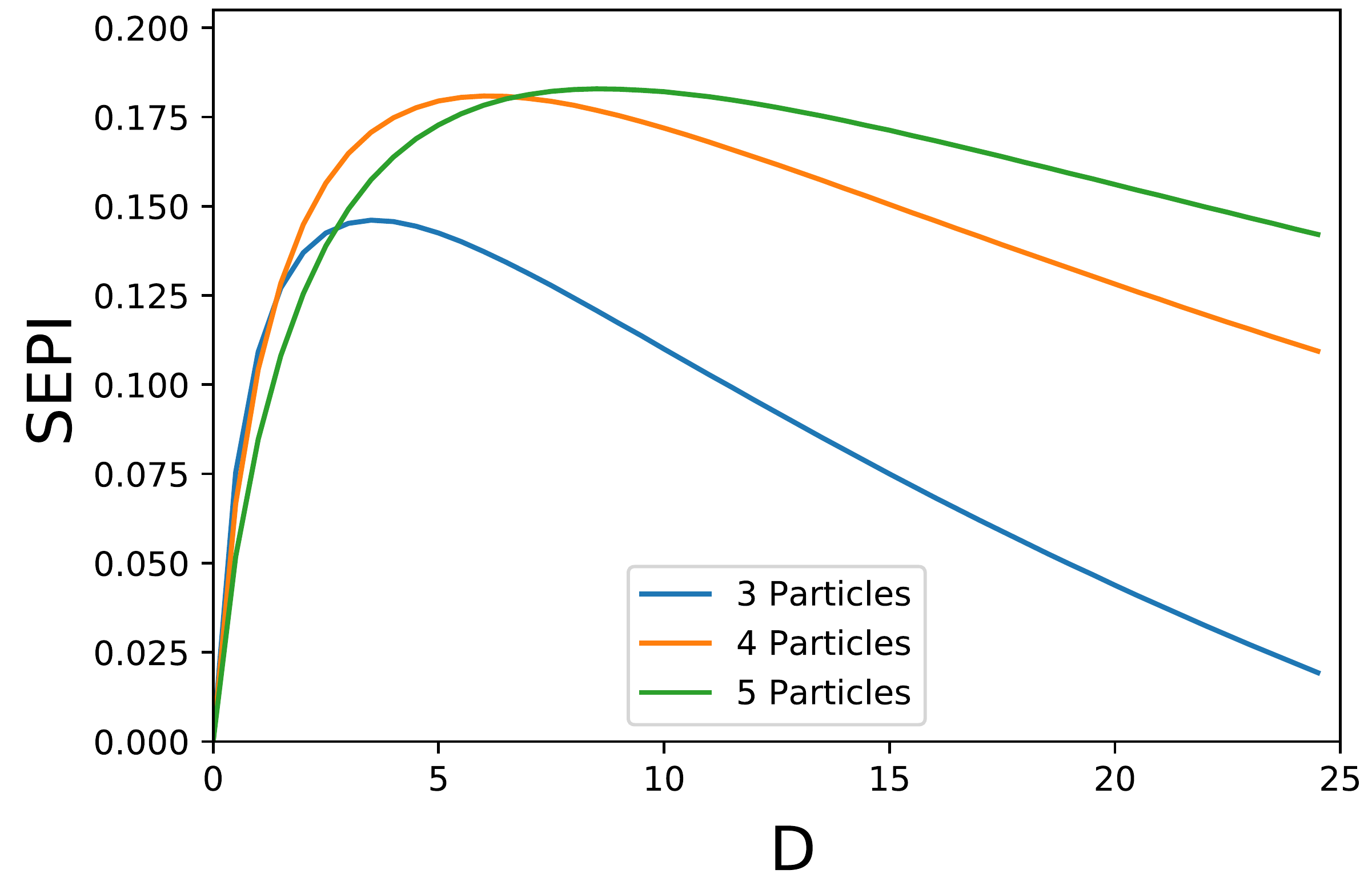}%{FIG1SIZEinset}
\caption{{\bf (a)}: SEP achieved for various system arrangements in a five-element network at the steady state for the configurations illustrated in Fig.~\ref{Config}. The curves associated with linear and non-critical cases are monotonically decreasing, offering no evidence of SEPI. {\bf (b)}: We show the behavior of SEP in a five-element maximally connected system against the dephasing rate $D$, highlighting the existence of a threshold value of $R$ after which ENAQT occurs. Panel {\bf (c)} shows the effect of increasing the system size upon ENAQT occurring in maximally connected networks. %(the inset magnifies the behavior of the curves in the region around $D\simeq{1})$. These were done by comparing the maximally connected scenario for each.
}
\label{ENAQTfig}
\end{figure*}

\subsection{No Ancillary System}
\label{noancilla}
%\textcolor{blue}{JAMES: please, make this part of the paper (till conclusions) more quantitative. Can we state an analytic condition for SEP? If not, can we give a plot of SEP against J? Throughout the paper: it's not clear which is the topology of the system being considered? where is the excitation placed at the beginning? where is the sink attached?}
In this Section, we consider a number of configurations in order to determine the requirements for ENAQT. We consider four archetypes as shown in Fig.~\ref{Config}: linear, loop, non-critical, and maximally connected. We use the term {\it path} or {\it pathway} to describe the route that an excitation might travel from the first qubit of a network (labeled as $1$) to the sink. In the linear configuration, the system {interacts} through nearest-neighbor couplings, which {gives} a single pathway from qubit $1$ to the sink [cf. Fig.~\ref{Config} {\bf (a)}]. {The} loop configuration [panel {\bf (b)}] also exhibits nearest-neighbor couplings {and} includes a direct interaction between the initial qubit and the final one. As will be explained later, such a direct connection is crucial for the phenomenology of ENAQT, and we thus dub it as {\it critical connection}. A configuration such as the one in Fig.~\ref{Config} {\bf (c)}, where the link between first and last element of the network only occurs through a set of non-nearest neighbor connections among the sites but lack of such a critical link will be referred to as non-critical. Finally, in the maximally connected configuration [Fig.~\ref{Config} {\bf (d)}] every site is coupled to each other. This offers the maximum number of pathways through the system.

In order to set a {benchmark}, we initially consider cold environments ($N_i=0$) and no ancilla coupled to ${\cal S}$. We {quantify} the relation between ENAQT and the properties of the system, and {find} that there are a number of conditions that our system should satisfy in order for ENAQT to occur. {To this end}, we introduce the parameter
\begin{equation}
R=\frac{J}{\gamma_\down},
\end{equation}
which quantifies the relative strength of the coherent coupling versus the incoherent environmental coupling for the damping of excitations into the local environments. Fig.~\ref{ENAQTfig} {\bf(b)} shows the effect of increasing $R$ values for a maximally interacting system of five qubits. ENAQT {only} takes place for sufficiently large values of $R$: while $R\lesssim1$ corresponds  to a monotonically decreasing behavior of SEP, a value of $R\simeq5$ already shows the existence of a region of values of $D$ where SEP increases with the dephasing rate [cf. Fig.~\ref{ENAQTfig} {\bf (b)}]. Such a region grows with $R$, until SEP assumes a monotonically increasing trend.

We have also addressed the dependence of ENAQT on the dimensionality of the system at hand [cf. Fig.\ref{ENAQTfig} {\bf (c)}]. In general, larger systems require a lower threshold in the value of $R$ for the enhancement effect to occur, and are associated with wider ENAQT ranges compared to smaller systems.  A larger system also displays a comparatively more significant effect. However, the range of beneficial dephasing values can grow at a greater rate than the magnitude of the effect and as a result, for certain $R$ and $D$ values, we can observe smaller systems getting a greater benefit than larger ones. %In the inset of Fig.~\ref{ENAQTfig} {\bf (c)}, for instance, for values of the dephasing rate up to $D\simeq1$, the $n=4$ case has larger SEP than the $n=5$ one. \ls{the difference is rather negligible: is the inset justified?} 
However, the maximum value of SEP will always increase with the size of the system. Our next observation addressed the various interaction pathways. We observed that non-critical and linear configurations, which both miss the direct connection between first and last element of the network, do not display any ENAQT effect, as SEP is a monotonically decreasing function of $D$. This {correlation between the existence of a direct link between first and last element of the network suggests a general} dependence of SEPI on the {network configuration} and thus, in turn, on the interference mechanisms that a chosen configuration entails. %We observed that systems with a greater number of interactions benefited more from ENAQT, for large enough values of $R$. %\textcolor{blue}{Mauro: please, James, explain better what is meant by "two system paths"..moreover, please use the right figure labelsWe also observed that there was a requirement for at least two system paths for the effect to occur. (Figure 1).} 

\subsection{Assessing the effect of quantum coherence}

We {now} address the role, if any, that quantum coherences set in the state of the network have in the settling of ENAQT. In order to provide a quantitative assessment, we use the $l_1$-norm of coherence proposed in Ref.~\cite{Coherence}, which reads
\begin{equation}
C_{l_1}(\rho)=\sum_{i,j}\left | \rho_{ij} \right |{-1},
\end{equation}
where $\rho_{ij}$ is a generic entry of the density matrix of the system and the double summation extends over all values of the indices $i$ and $j$.

\begin{figure*}[t!]
	\hskip1cm{\bf (a)}\hskip5.5cm{\bf (b)}\hskip5.5cm{\bf (c)}\\
	\includegraphics[width=.7\columnwidth]{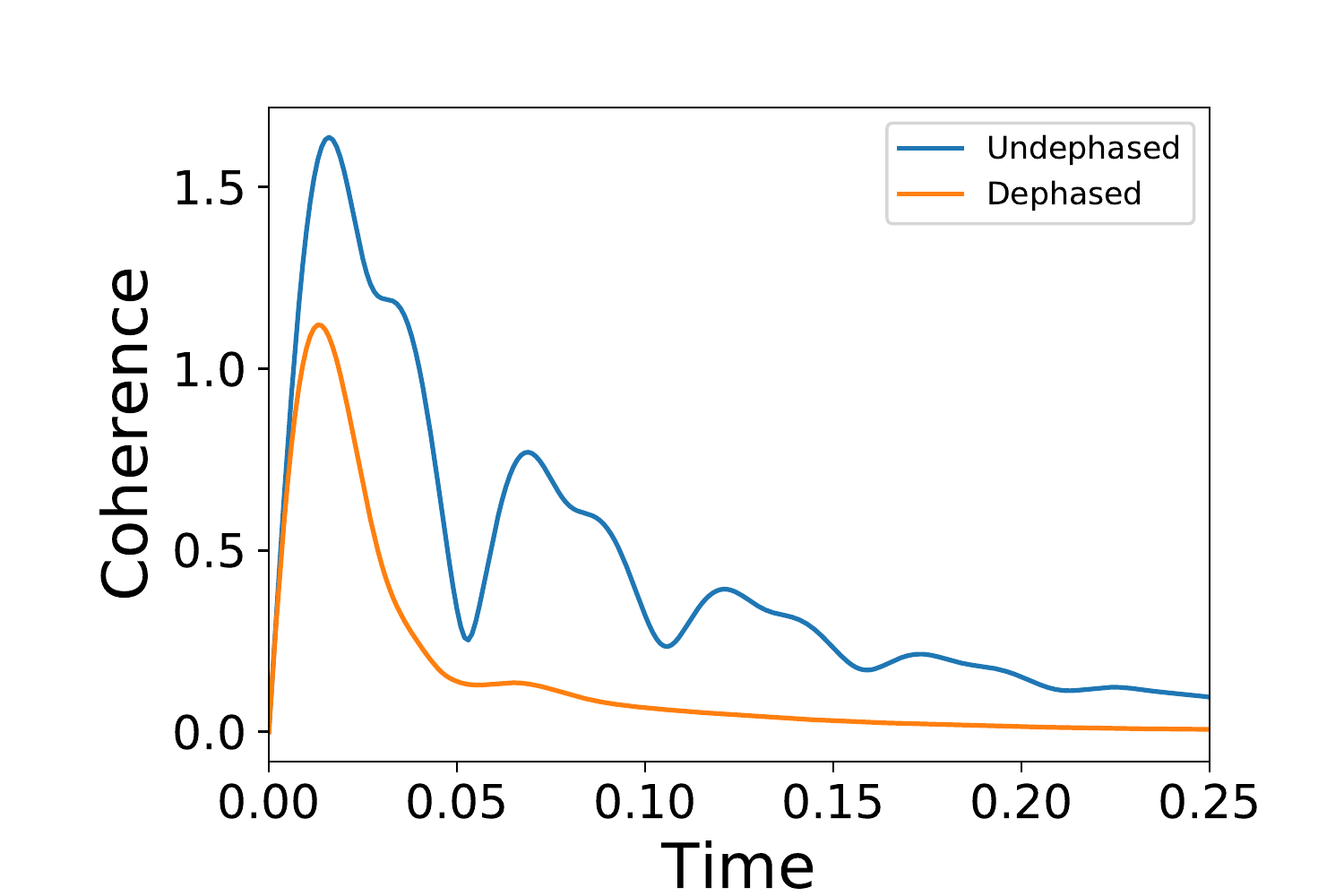}\includegraphics[width=.7\columnwidth]{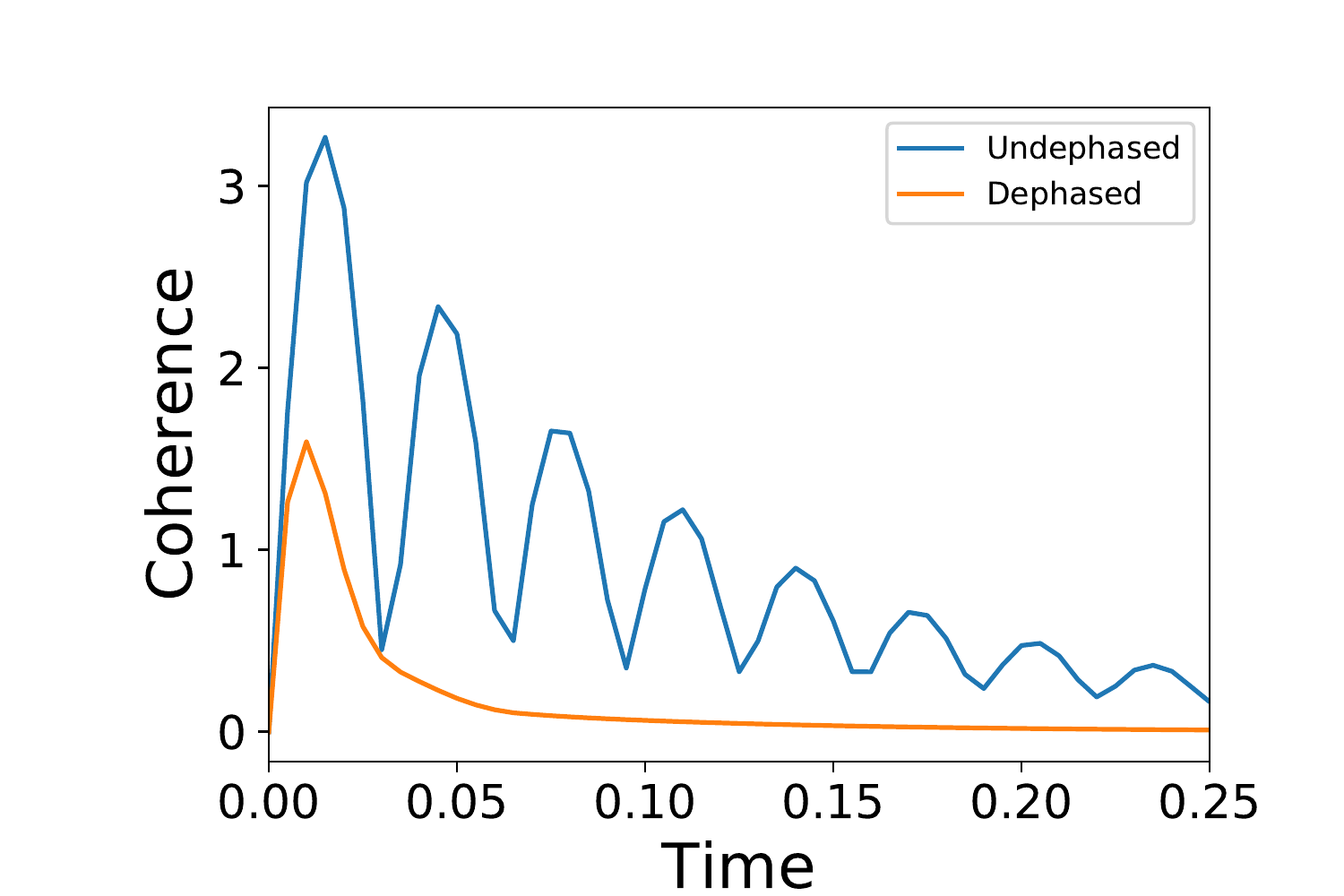}\includegraphics[width=.7\columnwidth]{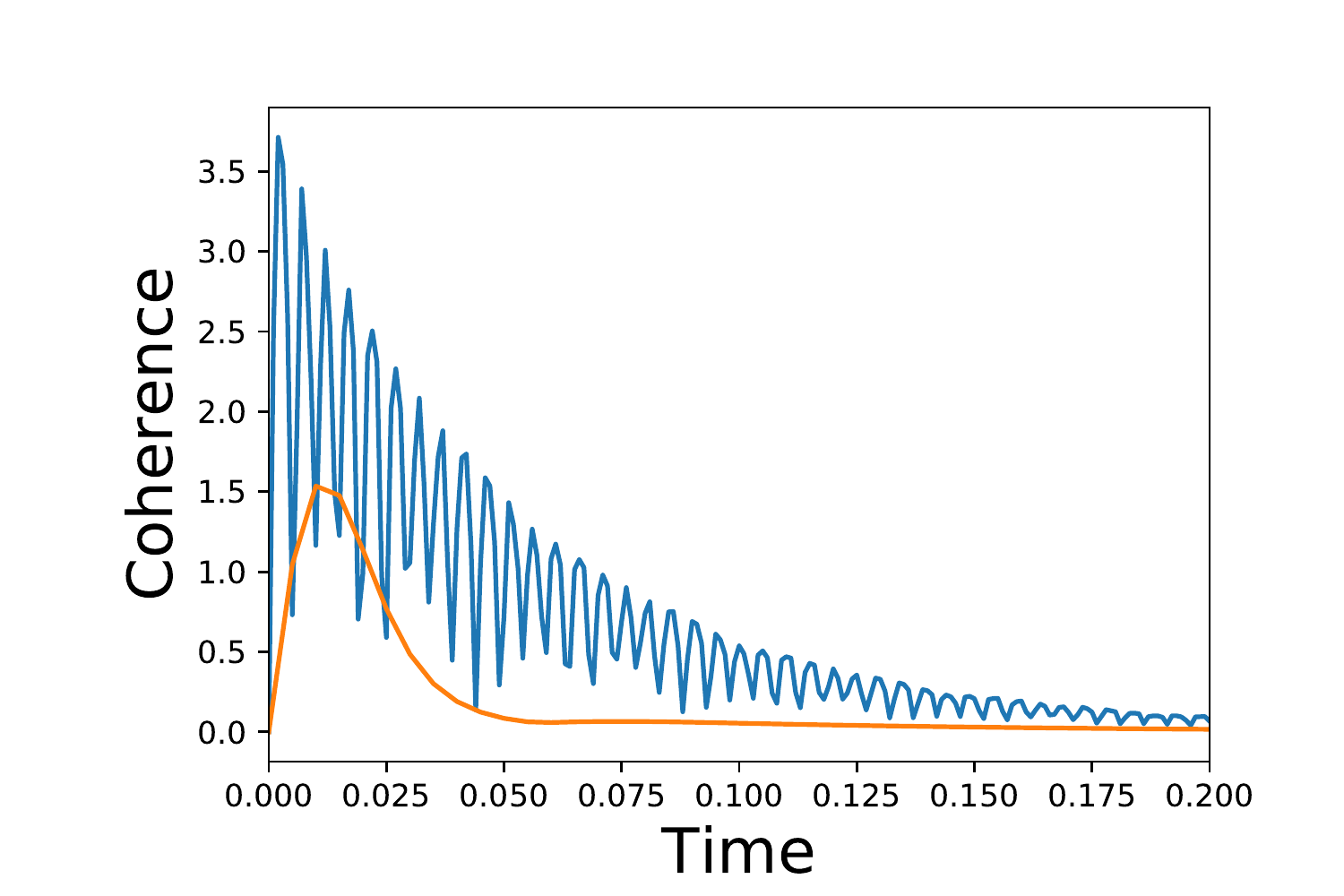}
	\caption{{Temporal} behavior of quantum coherences in the state of a maximally connected system of three {\bf (a)} and five {\bf (b)} qubits with $R=20$. Note that as system size increases, so does the coherence value. In both these panels, the curve showing the behavior of the dephasing affected system is obtained by using the value of $D$ at which SEP and ENAQT are maximum. Panel {\bf (c)} shows a five-qubit maximally non-critical configuration with $R=500$. The dephasing plot uses a dephasing strength that is optimal for the maximally connected configuration with $R=500$.}
	\label{coherences}
\end{figure*}

Fig.~\ref{coherences} shows the temporal behavior of $C_{l_1}(\rho)$, comparing the results corresponding to the dephasing-free and dephasing-affected dynamics, the latter calculated using the value of $R$ that optimize ENAQT.
Strong quantum coherences are associated with values of $R\gg1$, growing with the {network size}. While the undephased configuration shows a fading oscillatory behavior that however maintains a non-zero degree of coherences for a substantial time window, the dephased one displays a quick damping of coherences, which do not survive beyond the first period of oscillations of the $D=0$ case. However, despite the rapid disappearance of quantum coherence in the state of the system, Fig.~\ref{coherences} {\bf (a)} and {\bf (b)} are associated with the occurrence of ENAQT. On its own, this is significant evidence that quantum coherences do not appear to play a crucial role in the emergence of ENAQT. Moreover, the results displayed in Fig.~\ref{coherences} {\bf (c)} provide further useful information. Despite showing similar or larger degrees of coherence than in {\bf (a)} and  {\bf (b)}, panel {\bf (c)} does not showcases any ENAQT. As the configuration addressed there only lacks of the critical connection between the first and last qubit of the network, we conjecture the crucial role played by such a link in the establishment of ENAQT: quantum coherence alone are not sufficient for the effect to appear, but need to be complemented by the addition of a critical pathway between first and last element of the network.

\subsection{Assessing the Effect of Temperature}

\begin{figure}[b!]
	{\bf (a)}\\
	\includegraphics[width=.9\linewidth]{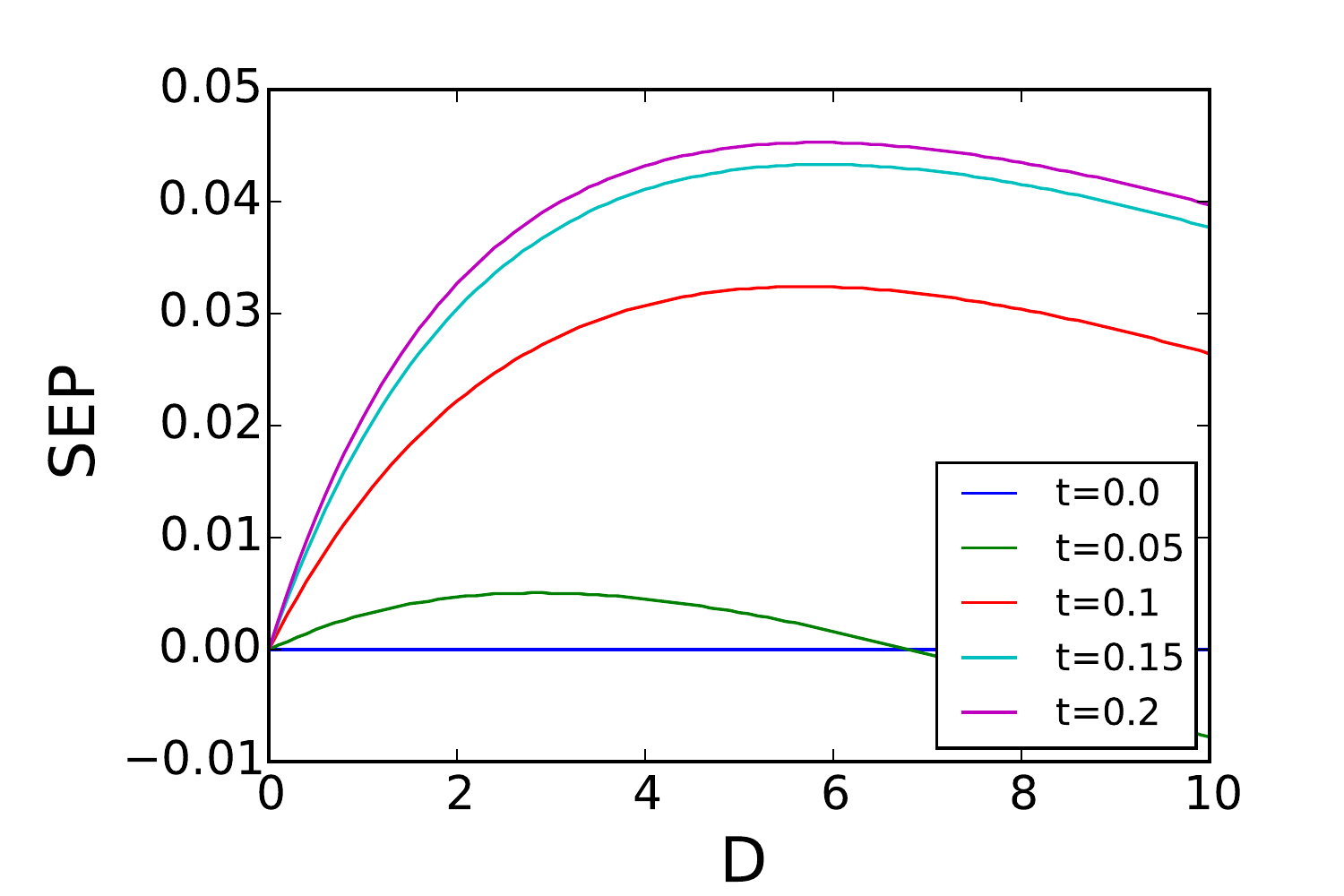}\\
	{\bf (b)}\\
	\includegraphics[width=.9\linewidth]{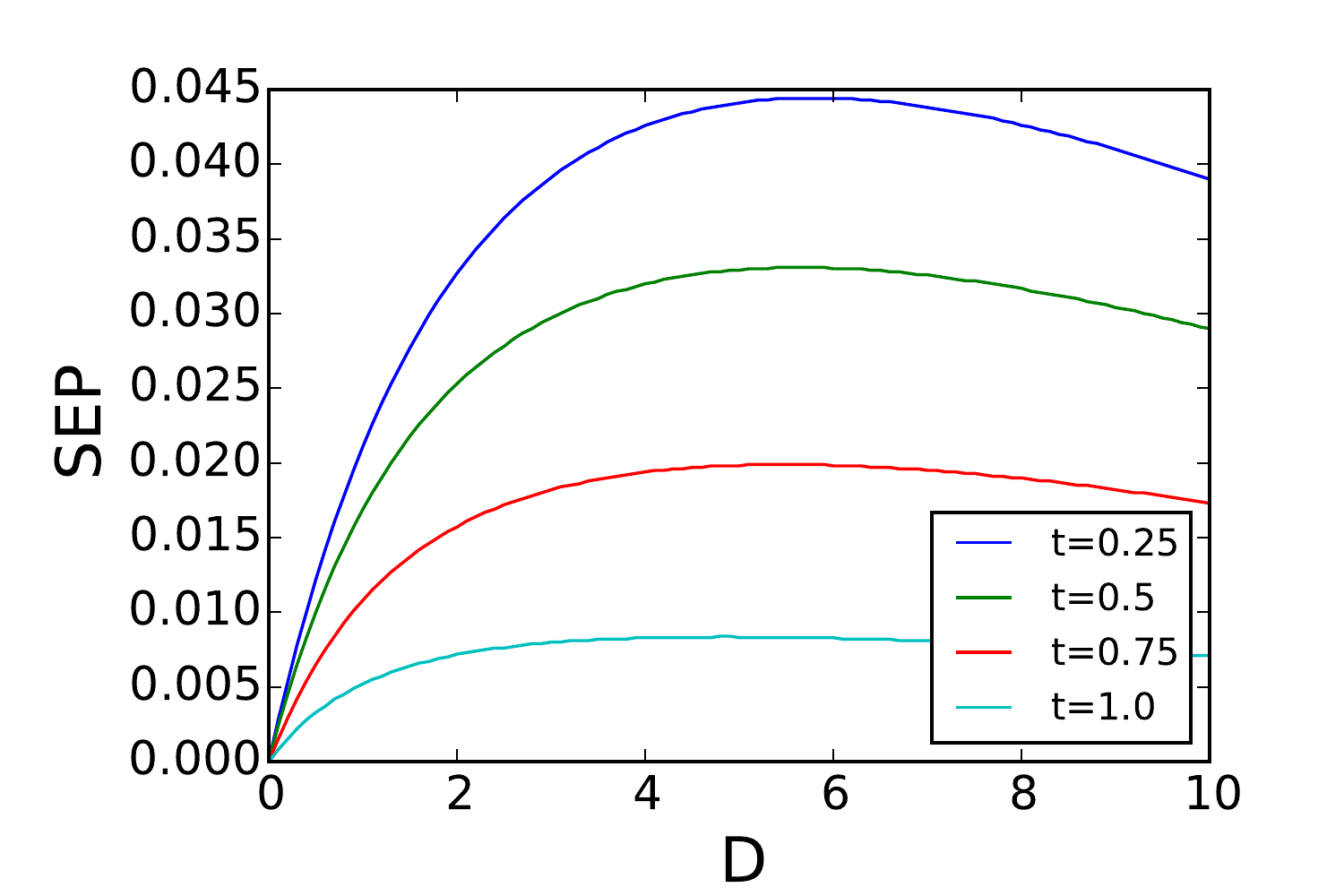}
	\caption{Assessment of temperature on the emergence of ENAQT. Panel {\bf (a)} shows the increasing ENAQT effect as time passes. Panel {\bf (b)} shows a waning ENAQT effect at longer times. %Neither of these plots include the steady state of the system as its steady state is one where {SEP}$=1$.
	The range of considered times is such that the steady-state is not reached. All the calculations presented are for a five-element maximally connected network and $N_1=1$ with all the other baths having no thermal excitation.}
	\label{dynamical}
\end{figure}

We now aim to characterize the influence that temperature has on the phenomenology of ENAQT. To {this end}, we {set} $N_i>0$. {Unlike} {the case with $T_i=0$}, we observe no sign of ENAQT at the steady state. However, such effect is present dynamically: snapshots of the evolution of the system at various dephasing strengths show that  ENAQT is present initially and disappears as we approach the steady state [cf. Fig.~\ref{dynamical}]. We observed the occurrence of such time-dependent enhancement for both  `directional' configurations where $N_1\neq0$ and all other $N_i=0$, and the case where $N_i\neq0$ for all the site in the network. However, the fully warm regime had a larger threshold value of $R$ for the occurrence of ENAQT than the directional regime. The addition of the ancillary system enriches the phenomenology of excitation-transport, as regimes where the effects of the warm environment are mediated by the ancillary system itself. However, we find common {features} throughout: as the number of `warm introduction sites' (i.e. sites exposed to a thermal environment additional excitations can enter) increases, the threshold value of $R$ required to observe ENAQT also increases. %Additionally, warm regimes that offer an excitation bypass around the system also have a larger minimum $R$ value. 

\subsection{Introducing the Ancillary System}
\label{incoherentancilla}

We now address the case where ${\cal A}$ is introduced in the overall system. In order to keep the computational effort to a reasonable level, but without affecting the generality of our conclusions, we have opted to consider a three-qubit system for this analysis. We shall first consider the case of incoherently coupled ancillary system [i.e. $Q=0$ in the Hamiltonian and $\Gamma_{\up,\down}\neq0$ in Eq.~\eqref{incoherentSA}]. Within this archetype, we study the excitation-transport performance under various dynamical regimes. Specifically, we considered $\Gamma_\up/\Gamma_\down=1$, for which there is no bias between the incoherent process that pulls excitations away from the system into the ancilla and the opposite process, $\Gamma_\up/\Gamma_\down>1$, where the ancilla supplies excitations to ${\cal S}$ at a rate larger than it takes them away, and the complementary situation of $\Gamma_\up/\Gamma_\down<1$, where depletion due to the system-ancilla interaction is strong. Both an isolated and open ancilla have been taken in consideration. In the first case, excitations could only enter or leave the ancillary system via its coupling to ${\cal S}$. This is not the case when an open ancilla is addressed, which significantly modifies the phenomenology of excitation-transport that we will highlight. 

\begin{figure*}[t!]
\hskip1.0cm{\bf (a)}\hskip4.0cm{\bf (b)}\hskip4.5cm{\bf (c)}\hskip4.0cm{\bf (d)}\\
\includegraphics[width=.26\linewidth]{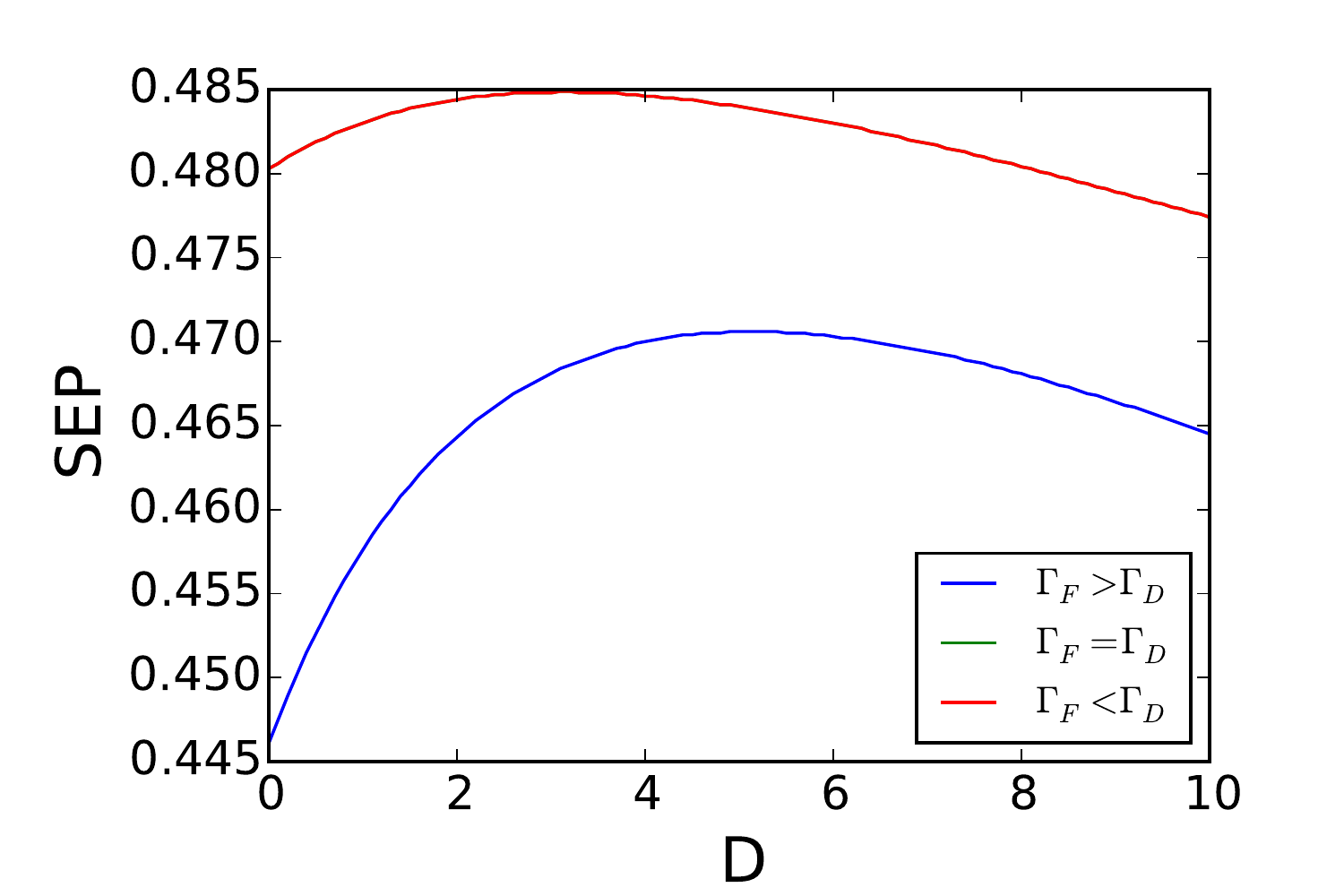}\includegraphics[width=.26\linewidth]{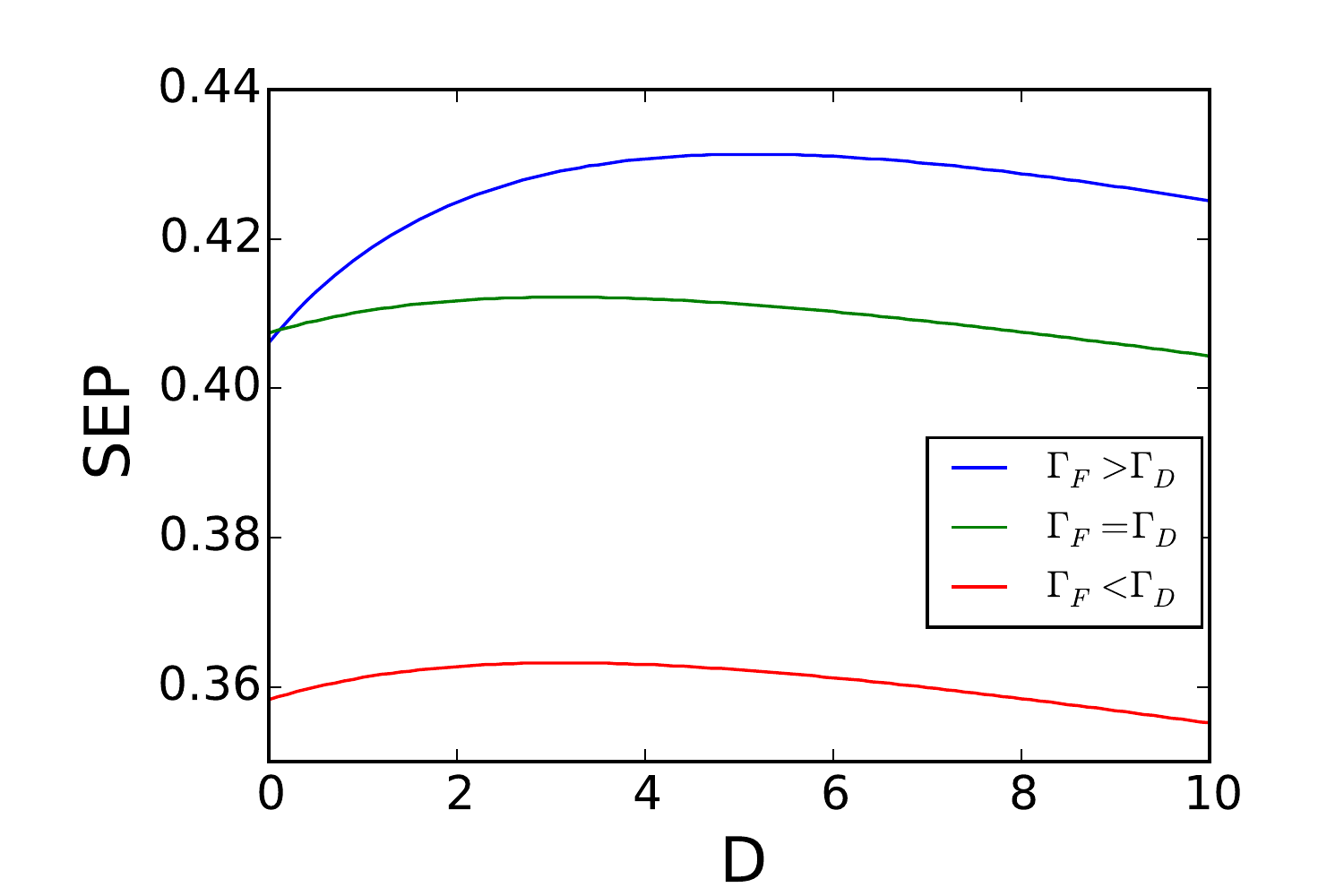}\includegraphics[width=.26\linewidth]{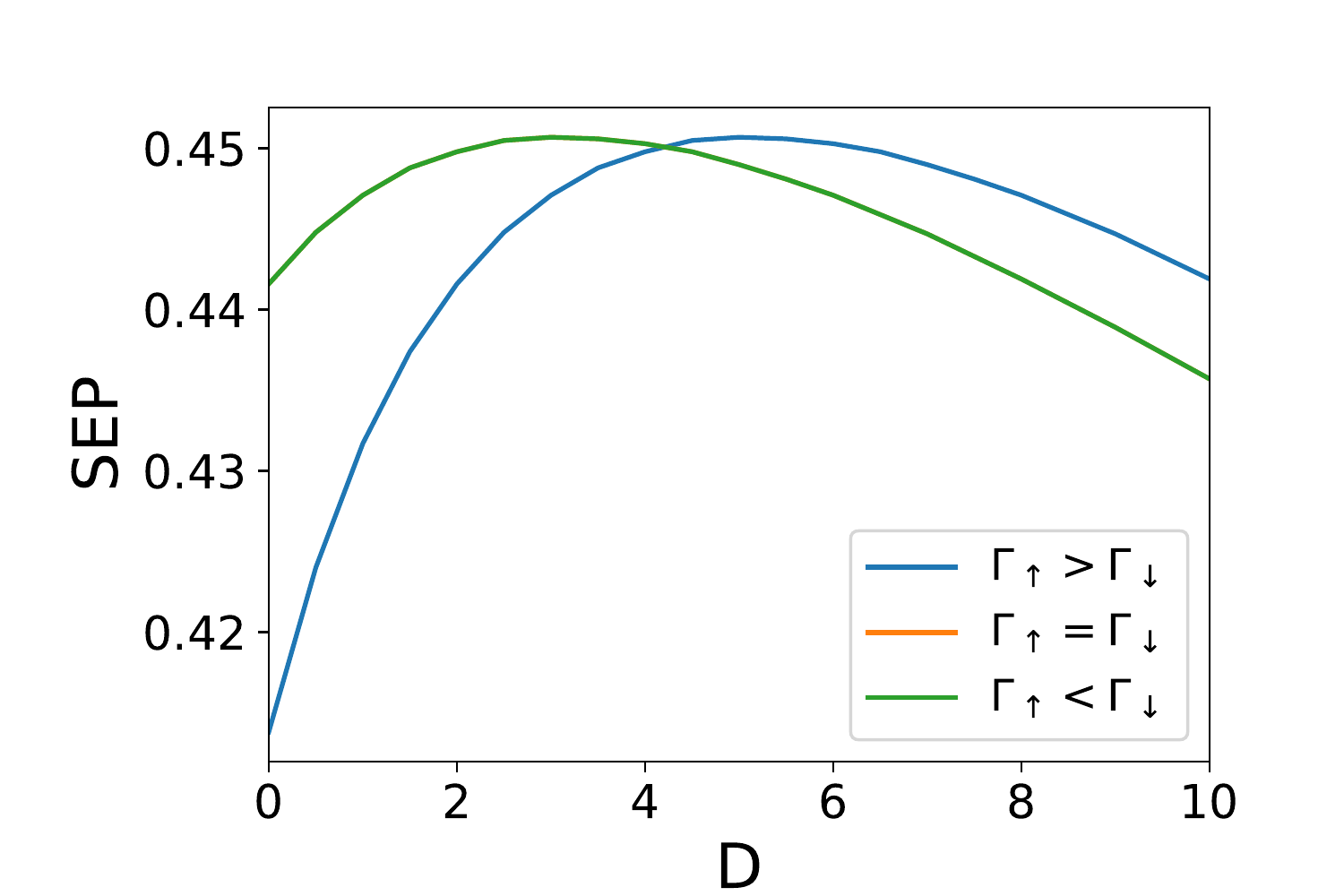}\includegraphics[width=.26\linewidth]{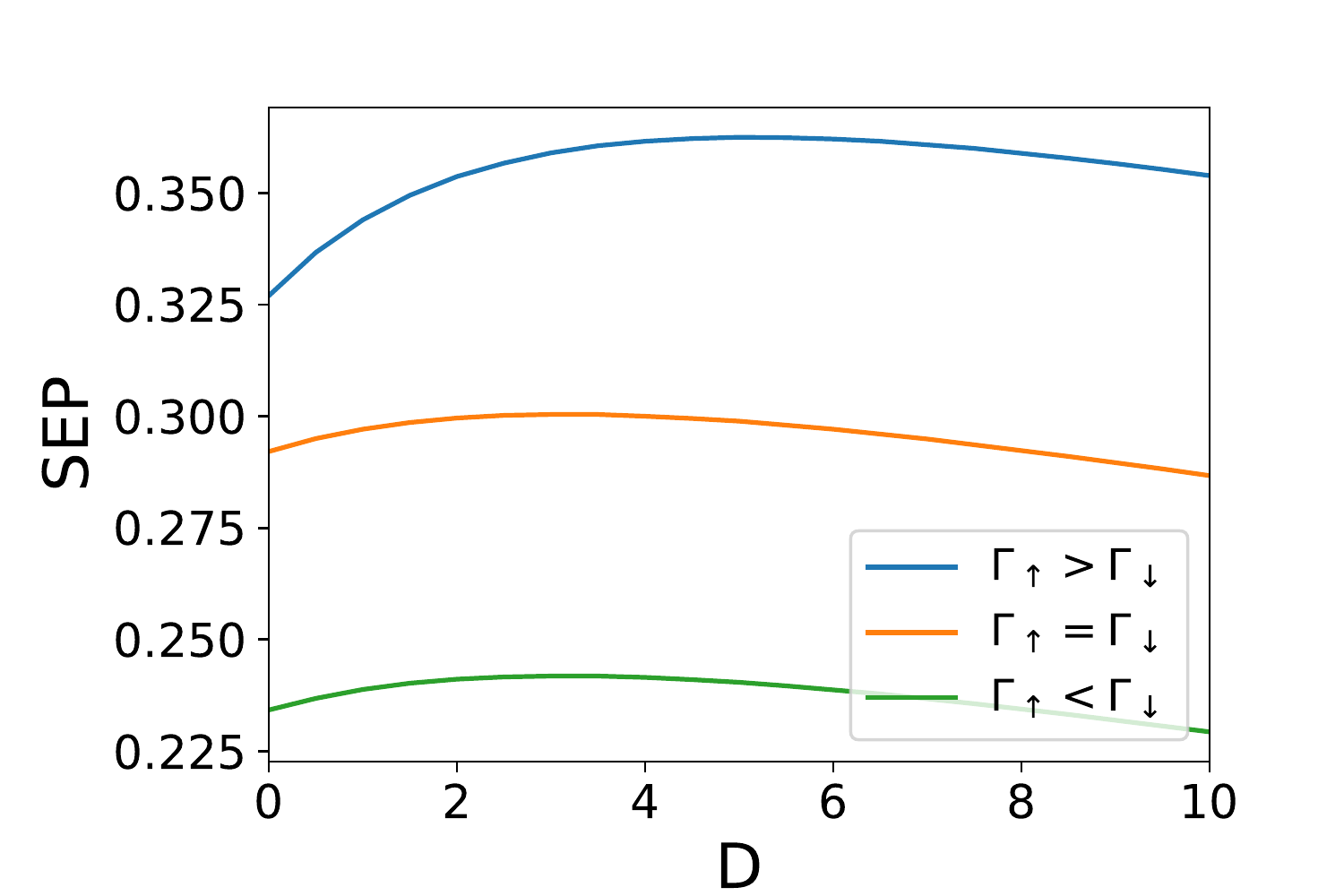}
\caption{SEP in a three-element maximally connected system for three different types of incoherent coupling: $\Gamma_\down=2\Gamma_\up$, $\Gamma_\down=\Gamma_\up$, and $\Gamma_\down=\Gamma_\up/2$. Panel {\bf (a)} shows the case of a communal ancilla without environment. Note that the curves corresponding to $\Gamma_\down=2\Gamma_\up$ and $\Gamma_\down=\Gamma_\up$ overlap. Panel {\bf (b)} shows the communal regime with environmental interface regime. Panel {\bf (c)} shows the individual ancilla without environment case (the curves corresponding to $\Gamma_\down=2\Gamma_\up$ and $\Gamma_\down=\Gamma_\up$ overlap). Finally, panel {\bf (d)} shows the case of individual ancillae with environment.}
\label{inco}
\end{figure*}

The corresponding results are summarized in Fig.~\ref{inco}. The incoherent coupling regime has a noticeably larger threshold value of $R$ to observe the ENAQT effect, when compared to the study reported in Subsec.~\ref{noancilla}. The maximum SEP for the individual isolated ancillary regime and the no-ancilla one turn out to be equal. We also observed that the maximum SEP for the communal isolated ancillary system was noticeably greater than both, though it had the least gain from ENAQT.

Next we consider the case of a coherently coupled ancillary system. We thus take $Q>0$ in Eq.~\eqref{hamiltonian} and assume $\Gamma_{\up,\down}=0$, so that Eq.~\eqref{masterequation} should be considered to model the dynamics. As done above, we consider both closed and open ancillary systems. %Unlike the incoherent archetype, we only considered one type of coherent coupling, namely the Heisenberg XXX coupling which is also present within the system. 
When {compared} against the occurrence of ENAQT, this {configuration} results in a threshold value of $R$ that is lower than the one for both the incoherent ${\cal S}-{\cal A}$ coupling the case with no ancilla at all. This is very interesting, as it implies that the interaction with an ancilla {promotes} the occurrence of ENAQT. The effect can be understood as analogous to the observed reduction of the threshold value of $R$ when larger (coherently coupled) systems are studied: {\it de facto}, in this situation, ${\cal A}$ is akin to an extension of ${\cal S}$. 

\begin{figure}[b!]
{\bf (a)}\\
\includegraphics[width=\linewidth]{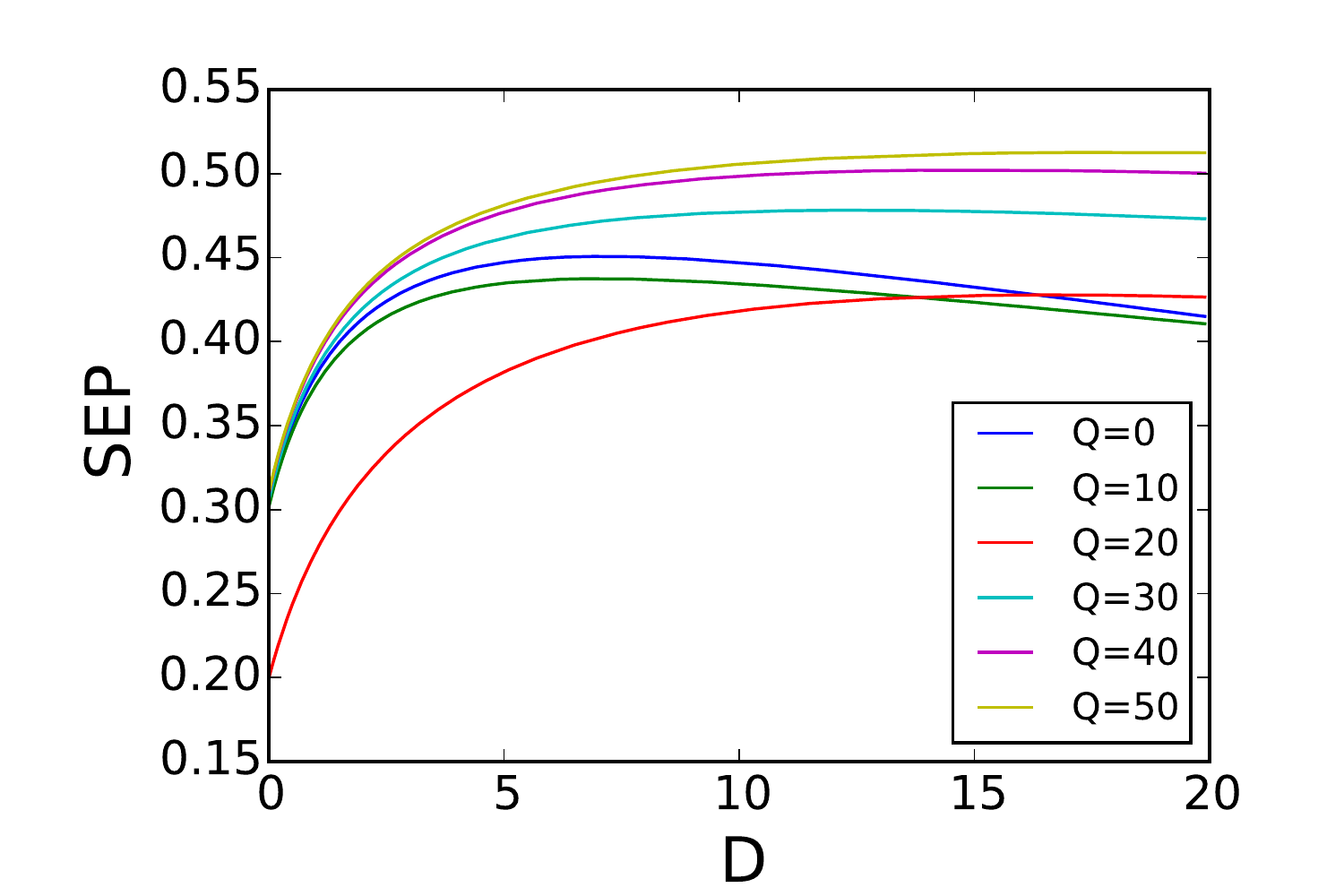}\\
{\bf (b)}
\includegraphics[width=\linewidth]{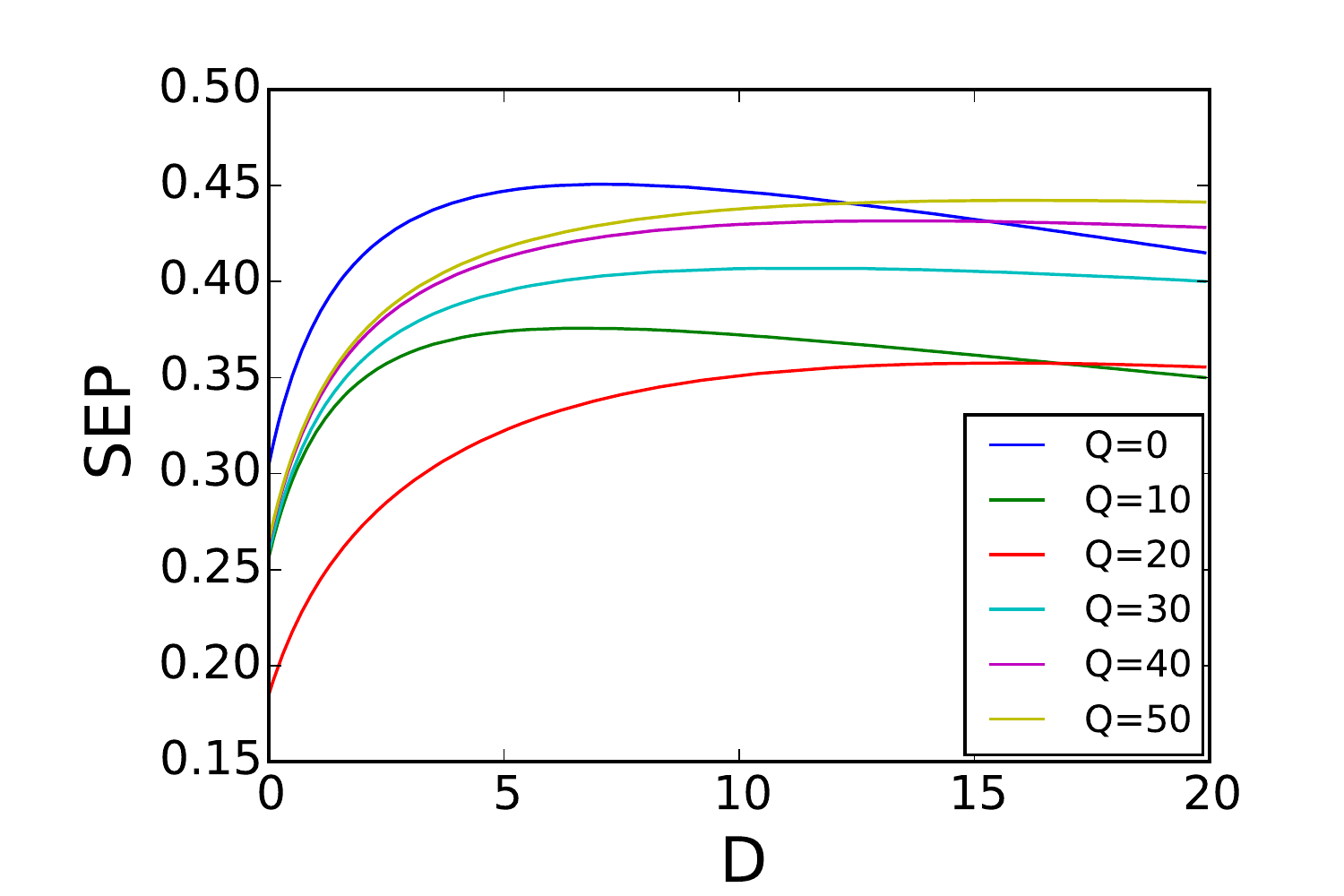}\\
%{\bf (c)}\\
%\includegraphics[width=.9\linewidth]{FIG3OFFIND}\\
%{\bf (d)}\\
%\includegraphics[width=.9\linewidth]{FIG3ONIND}\\
\caption{Panels {\bf (a)} and {\bf (b)} show plots for the communal coherently coupled ancilla regime where $R=20$. Panel {\bf (a)}  is the regime where the ancilla is isolated and does not interact with an environment. Panel {\bf (b)} is the regime where the ancilla is interacting with an environment.}
\label{coherent}
\end{figure}

The extra degree of freedom provided by the possibility of tuning the value of the system-ancilla coupling rate $Q$ allows for the characterization of ENAQT and SEP against variation of such a parameter. This is done in Fig.~\ref{coherent}, where we have taken $R=20$. The ENAQT effect is highly dependent on the relative value of $Q$ with respect to $R$. When $Q\approx R$, we observe a drop in SEP, but an increase in the ENAQT effect. We see the best ENAQT when this occurs as the ancillary and qubit system were effectively behaving as a slightly larger qubit system. Additionally, we saw that when $Q>R$, SEP grew. For the isolated ancillary system, configurations with large $Q$ performed best, in terms of SEP, as excitations were more likely to transfer through the ancilla. {Was} the ancillary system no longer isolated, it would represent another source for excitation loss. {In this configuration, the excitation transfer would perform as in the $Q=0$ case.}

In this regime, we have evidence of non-Markovian dynamics. In order to {quantify the degree of non-Markovianity, we use an} instrument put forward in Ref.~\cite{BPL, review} and based on the trace distance, {which is} defined as 
\begin{equation}\label{td}
  D(\rho_1(t),\rho_2(t))=\frac{1}{2}||\rho_1(t),\rho_2(t)||_1,
\end{equation}
where $||\cdot||_1$ is the trace norm and $\rho_{1,2}(t)$ are two density matrices of the same system. The trace distance is null (equal to one) for indistinguishable (fully distinguishable) quantum states. It is contractive under Markovian dynamics, meaning that the trace distance between two completely distinguishable initial states of a system exposed to a Markovian environment is {a non}-increasing quantity. That is, the rate of change of the trace distance will always be negative. Any deviation from such behavior, has to be associated with an evolution that is non-Markovian in nature, for instance due to {the back-action of the environment}. Based on such considerations, it is possible to define a measure of non-Markovianity as follows \cite{BPL}
\begin{equation}\label{blp}
  \mathcal{N}=\max\int_{dD/dt>0}\frac{dD}{dt}dt,
\end{equation}
where the maximization is made over all possible pairs of initial states of the system. While observing a temporary increase in the trace distance is sufficient to conclude that the dynamical map is non-Markovian, the converse is not true: non-Markovian evolutions might exist such that the trace distance decreases monotonically. In this sense, the condition $dD/dt>0$ is only a witness for non-Markovianity.

\begin{figure}[b!]
{\bf (a)}
	\includegraphics[width=\linewidth]{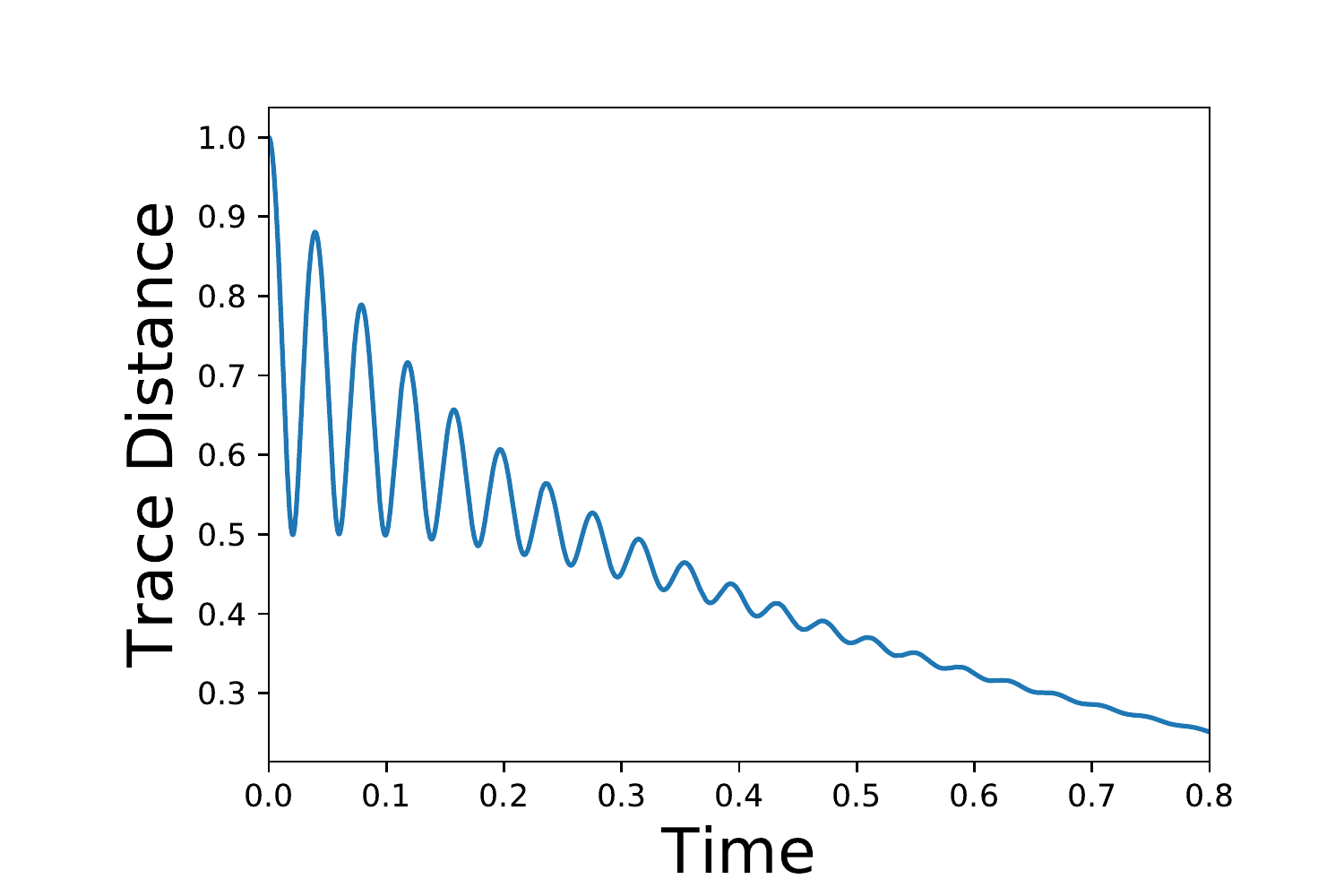}\\
{\bf (b)}
	\includegraphics[width=0.9\linewidth]{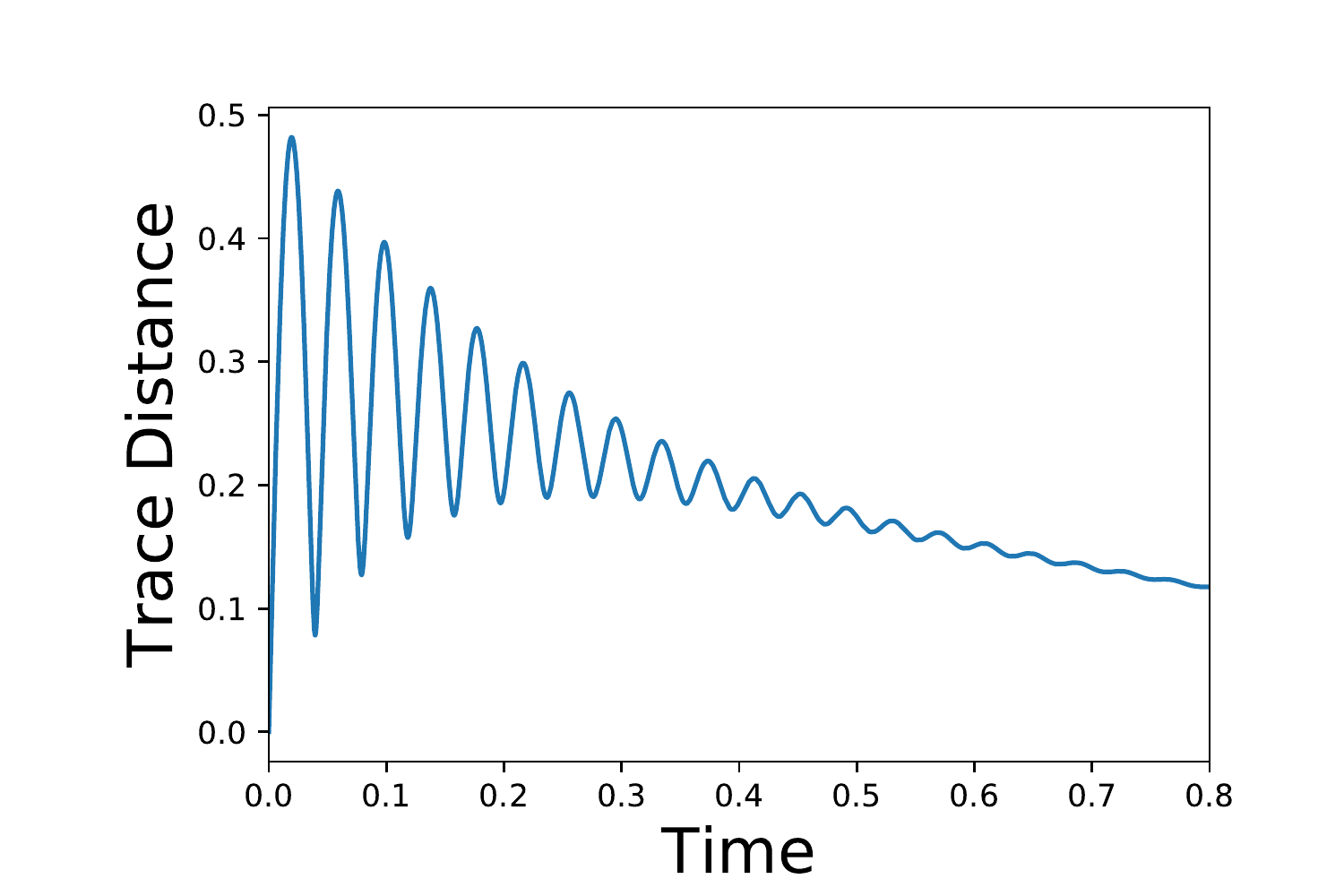}
	\caption{Panels {\bf (a)} and {\bf (b)}  show the time behavior of the trace distance for the communal coherently coupled ancilla regime where $Q \approx R$. In both panels, the trace distance is measured on the ancilla. Panel {\bf (a)} shows the scenario where the perpendicular states begin on the ancilla. This allows us to observe larger trace distance values, allowing for easier detection of non-Markovianity. Panel {\bf (b)} shows the  scenario where the perpendicular states are within the qubit system. While this scenario reduces the ability to detect non-Markovianity, it is the more realistic scenario. In both plots, we see revivals which suggests that there is non-Markovian behavior occurring.}%assessed the regime where the states of the system-ancilla are perpendicular with respect to the ancilla. This regime is more effective for detecting non-Markovianity through ancilla measurement.{\bf (b)} is the regime where the states are perpendicular with respect to the qubit system. This regime is more accurate to our scenario. Both plots show many revivals which is a marker for non-Markovianity.
\label{nonMarkPlot}
\end{figure}

In our approach, rather than considering the state of the system itself, which is multipartite, we assess non-Markovianity from the perspective of the ancillary system. {This choice greatly simplify the calculations}. We also argue that, if non-Markovian dynamics is observed in the ancillary system due to their coupling with the primary system, then, {by symmetry, the primary system dynamics will be also non-Markovian.} Fig.~\ref{nonMarkPlot} shows our results, revealing that a strong non-Markovian behavior is indeed witnessed in the dynamical system studied here. The non-Markovian environmental mechanism acts just as an additional coherent resource. 
{In fact, the enhancement can be equivalently obtained by} a suitably greater value of $R$, i.e. without the introduction of qualitatively significant new features. %As such, non-Markovian systems are useful to encourage these effects but do not stand out. \ls{This sentence is not clear and not quantitative}

\begin{figure}[t!]
	\includegraphics[width=\linewidth]{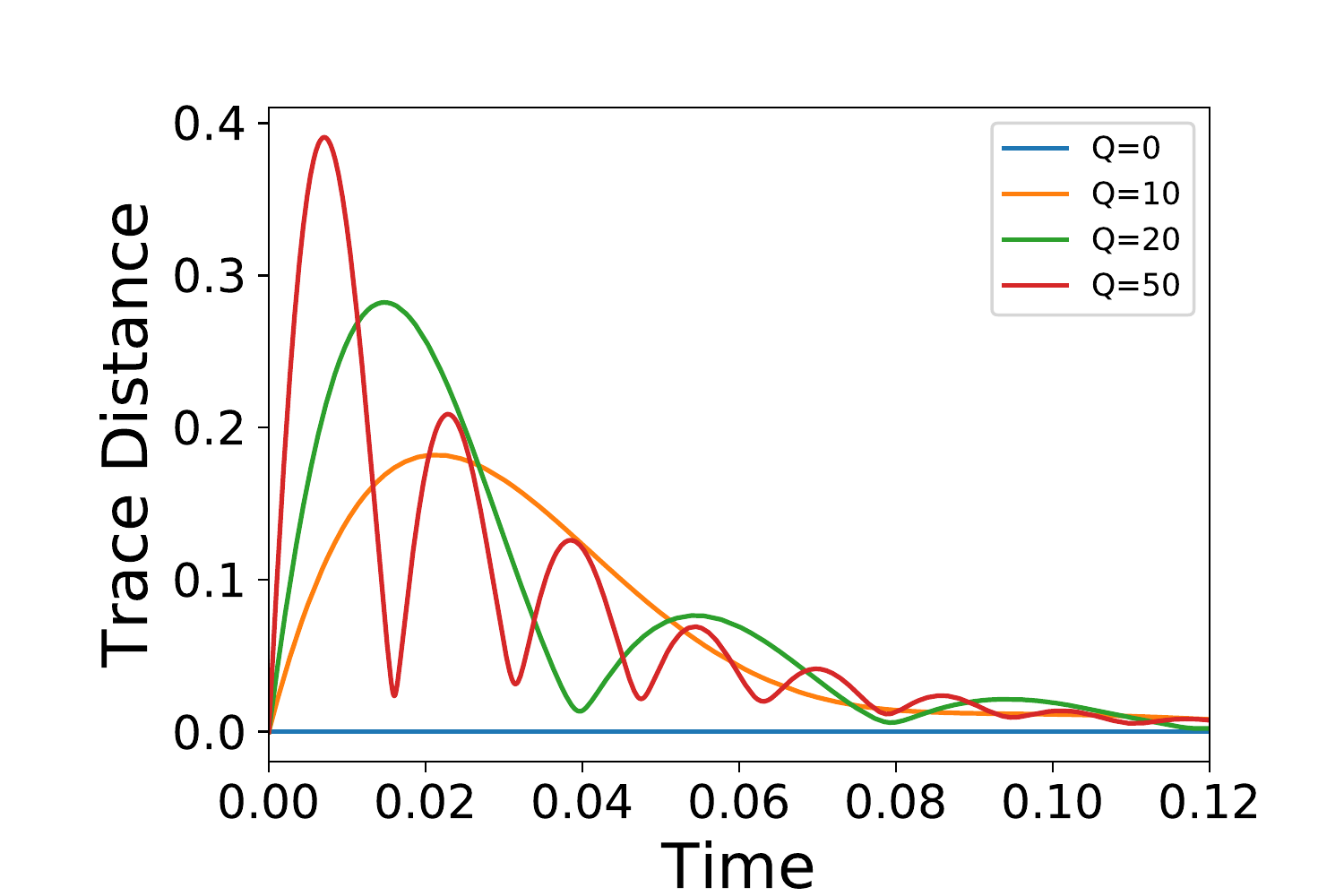}
	\centering
	\caption{This plot shows the trace distance for the three particle system coupled both coherently and incoherently with an ancillary qubit where the incoherent coupling is $\Gamma_\down=\Gamma_\up=1.0$ and $R=20$. Note that the revivals become more frequent and with greater magnitude as $Q$ increases in magnitude.  }
	\label{coheincohe}
\end{figure}

The {previous} analysis is complemented by the study of the mixed-coupling case corresponding to $Q, \Gamma_{\up,\down}\neq0$ [cf. Fig.~\ref{coheincohe}]. As observed in previous regimes, the inclusion of the ancillary-environment interface causes a noticeable reduction in SEP as well as a reduction in the ENAQT effect. When $Q\neq0$, we observe non-Markovian dynamics dependent upon the relative strengths of the incoherent couplings and the environmental coupling strengths.

\begin{figure}[b]
{\bf (a)}\hskip4cm{\bf (b)}\\
\includegraphics[width=.5\linewidth]{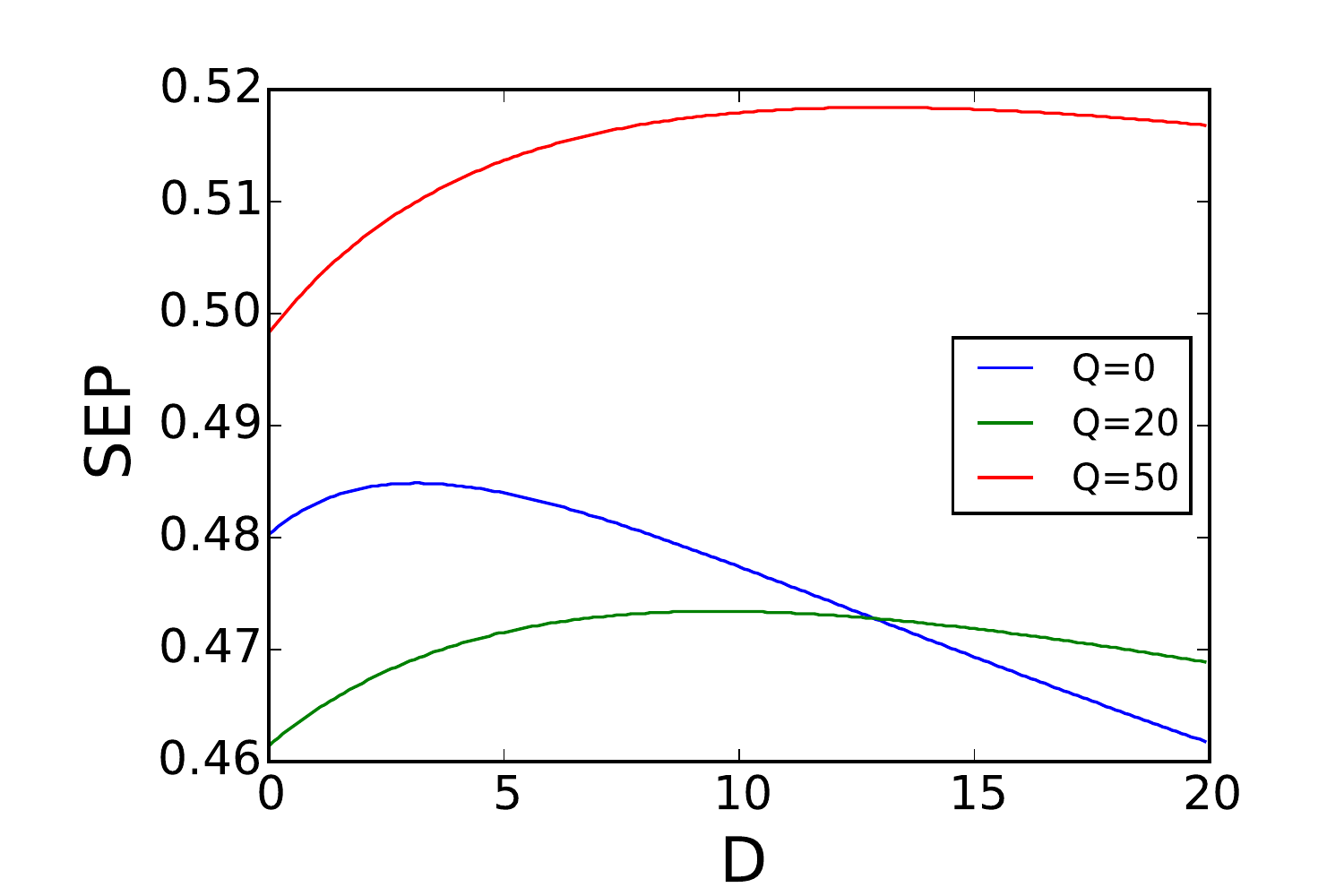}\includegraphics[width=.5\linewidth]{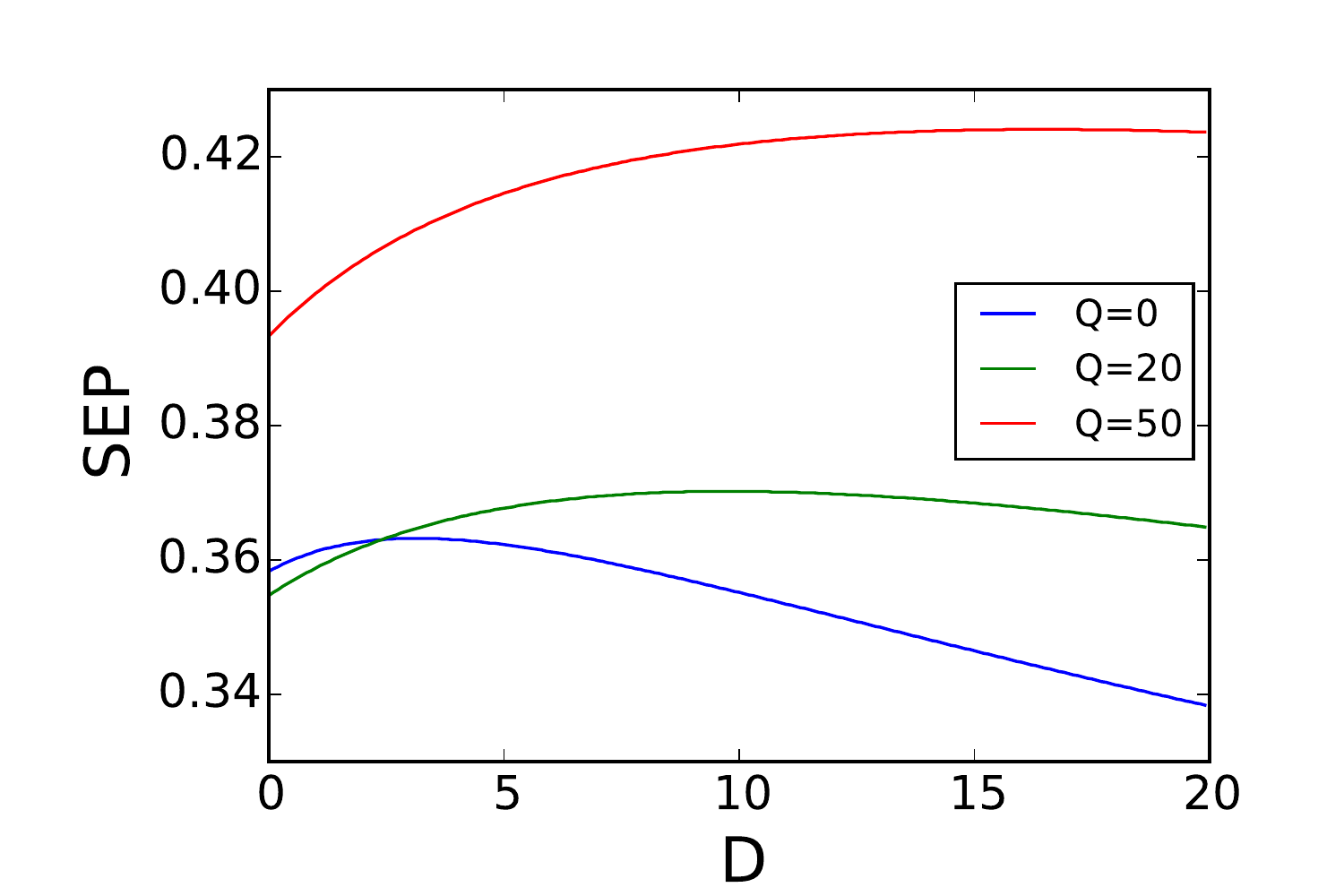}\\
{\bf (c)}\hskip4cm{\bf (d)}\\
\includegraphics[width=.5\linewidth]{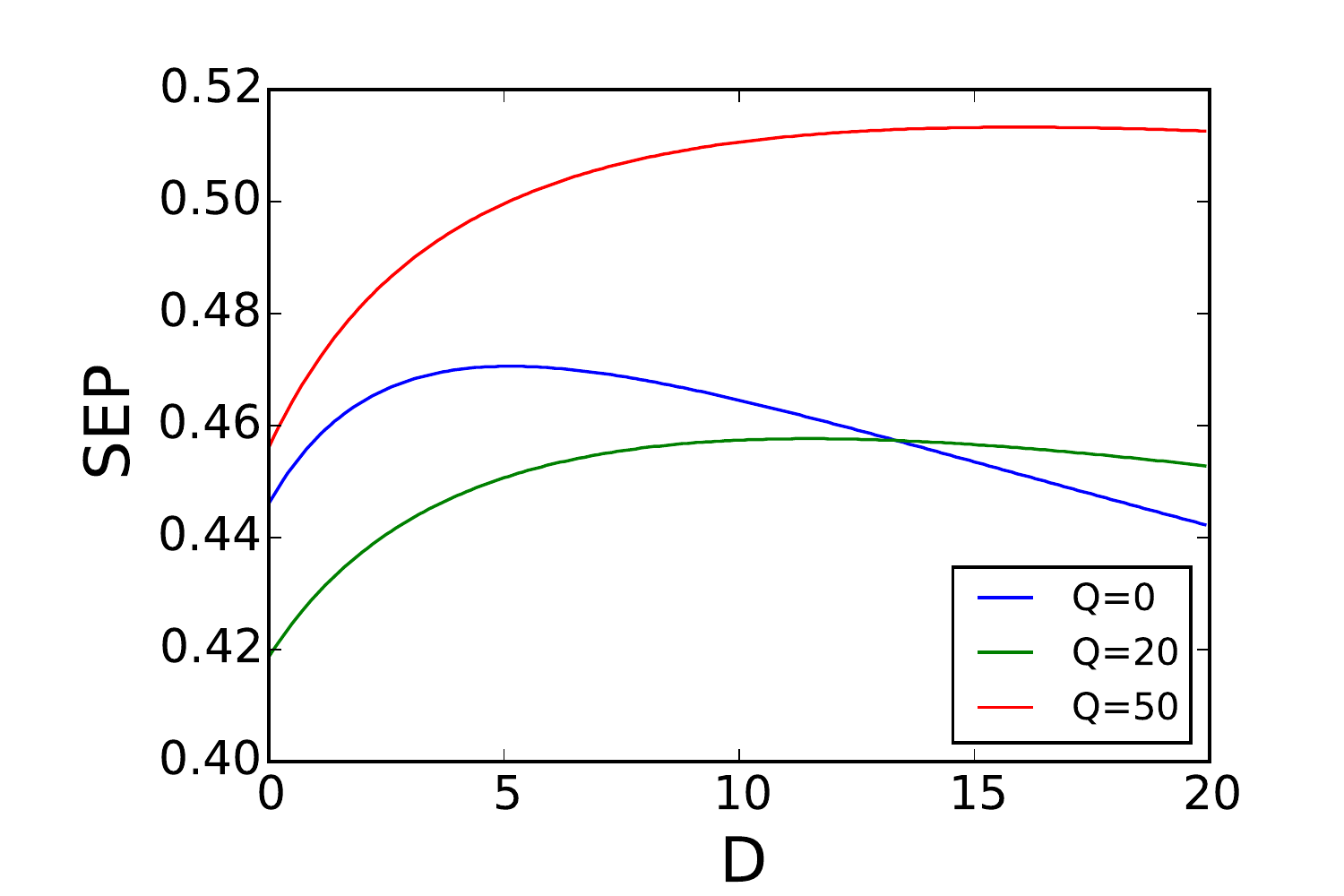}\includegraphics[width=.5\linewidth]{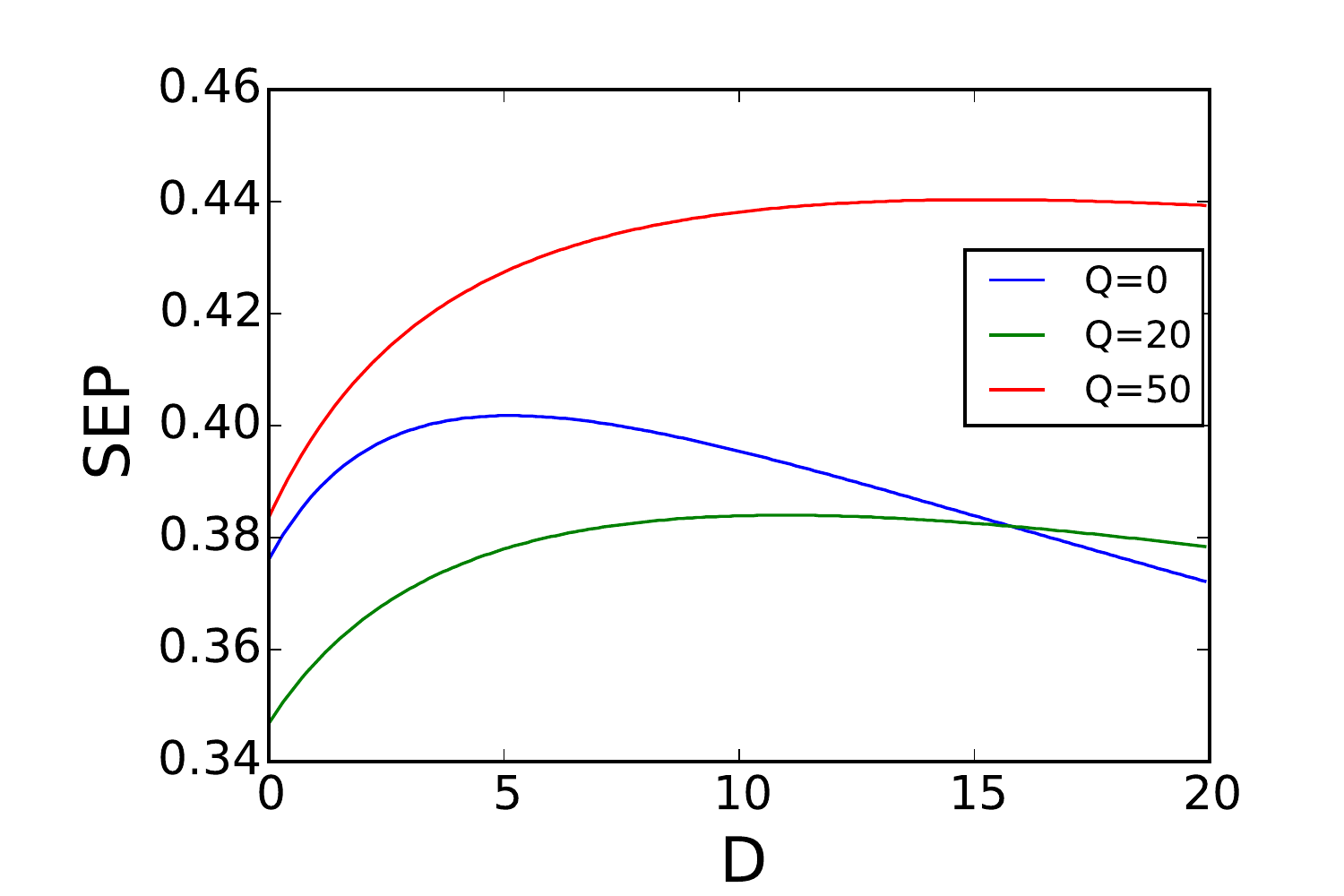} \\
{\bf (e)}\hskip4cm{\bf (f)}\\
\includegraphics[width=.5\linewidth]{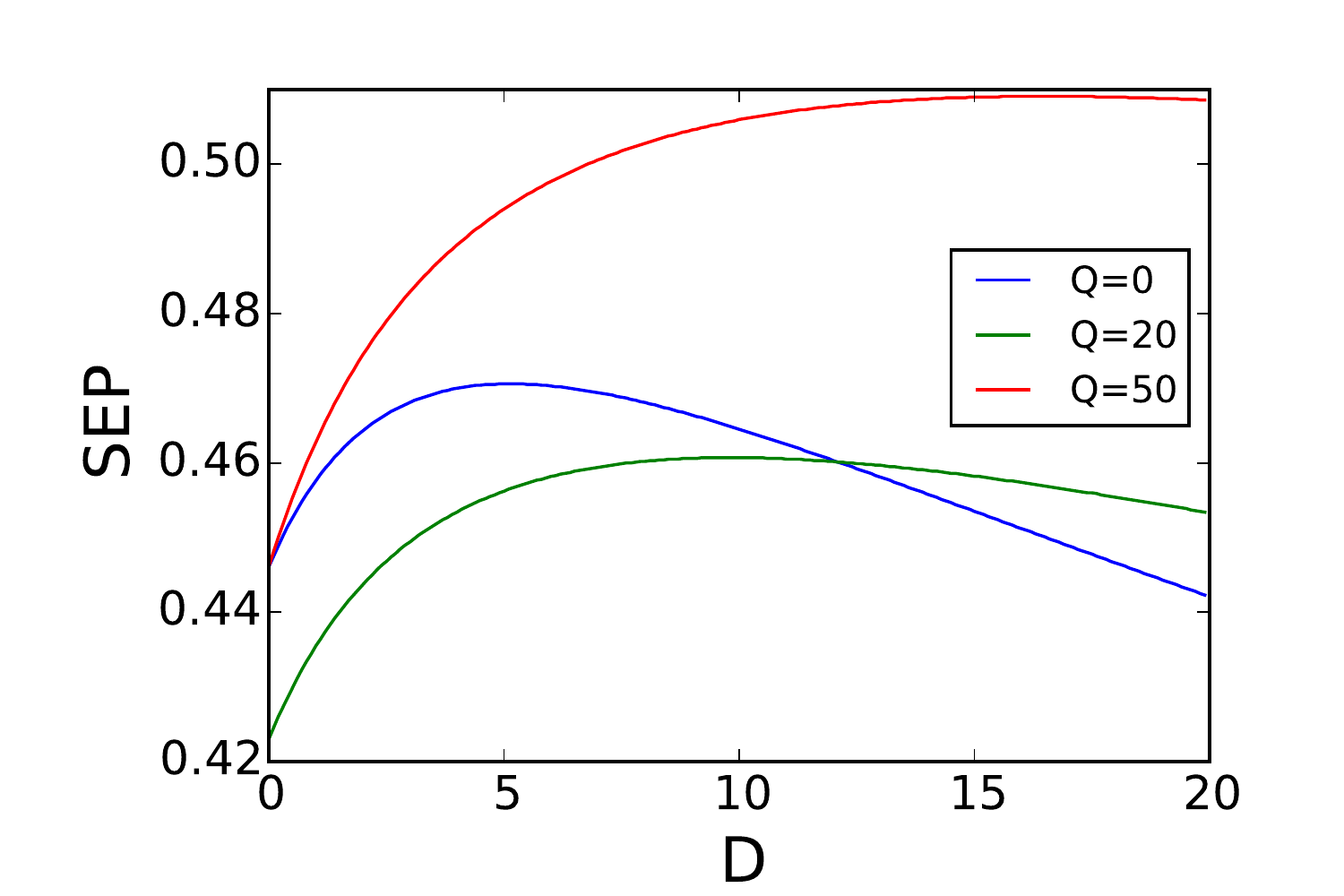}\includegraphics[width=.5\linewidth]{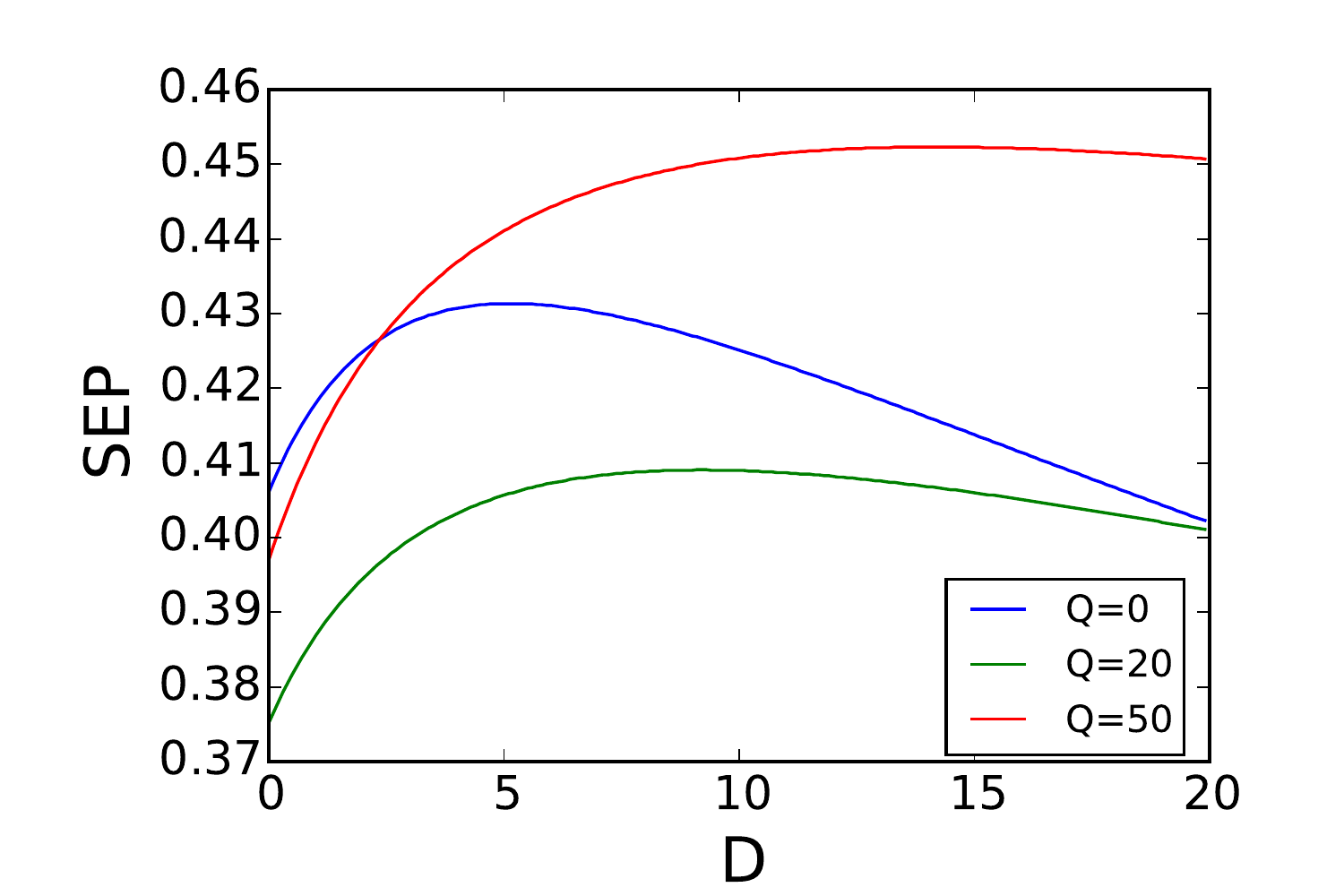}\\
%\includegraphics[width=.9\linewidth]{FIG4DISINDOFF}\\
%{\bf (d)}\\
%\includegraphics[width=.9\linewidth]{FIG4DISINDON}
\caption{These plots show the mixed regime where the incoherent component favors dissipation [i.e. $\Gamma_\down>\Gamma_\up$] [panels {\bf (a)} and {\bf (b)}]; does not favor either dissipation or feedback (i.e. $\Gamma_D=\Gamma_F$) [panels {\bf (c)} and {\bf (d)}]; and where the incoherent component favors the feedback (i.e. $\Gamma_D<\Gamma_F$) [panels {\bf (e)} and {\bf (f)}]. Several coherent coupling strengths are considered and the ancillary-environment interface is either excluded [{\bf (a)}, {\bf (c)} and {\bf (e)}], or included [{\bf (b)}, {\bf (d)}, and {\bf (f)}].}
\label{mixed}
\end{figure}

For incoherent coupling strengths that are large with respect to the coherent couplings [i.e. for $\Gamma_{\up,\down}\gg Q$], we see very limited or possibly no non-Markovianity. Large relative coherent couplings ($Q\gg\Gamma_{\up,\down}$) are associated instead with pronounced non-Markovian features. In the mixed regime, we observed that the $R$ requirement was dependent on this relative strength as well. We found that as $Q$ became the more dominant {parameter} that our $R$ requirement reduced [cf. Fig.~\ref{mixed}].%When $Q$ is non-zero, we observe some degree of non-markovian dynamics with its lifetime being dependent upon the relative strengths of the incoherent couplings and the environmental coupling strengths (Appendix C). As observed in previous regimes, the inclusion of the ancillary-environment interface causes a noticeable reduction in SEP. For the communal ancillary regime, we see that 

\section{Discussions and conclusions}
\label{conclusions}

We have addressed the question of the emergence of environment-induced advantages in the transfer of excitations across a network of interacting particles, studying explicitly the effects that different geometries of the network have on the efficiency of excitation-transport. 

Our chosen figure of merit for the efficiency of excitation-transfer, namely {the sink excitation probability (SEP)}, exhibits dephasing-induced enhancement with respect to the fully unitary case, but only if a suitable {network configuration} is ensured. In particular, we have highlighted the critical role played by the availability of a direct {link} connecting the sending and receiving sites in the network, which interferes constructively with the paths going through the other sites owing to local dephasing noise of modest entity. %Such links ensure the emergence of interfering paths along which excitations are transferred. The pattern of mutual interference among such paths appears to be strongly affected by the presence of a local dephasing mechanism.  
% With these two conditions, an interference pattern is created through the system which restricts the mobility of excitations through the system. This is resolved by the action of decoherence and the additional requirements we observed for ENAQT to occur are due to competing decoherences. We can easily oberseve this for environmental dissipation and the incoherent ancillary coupling restricting the availability of ENAQT. However, not all decoherences are equal. In the case of the environmental dissipation, the decoherence is also destructive with respect to excitation transfer. If we consider excitation transfer as a game, non-destructive coherences are competetition whereas destructive coherences are trying to stop the game altogether.
%Our Markovian environment had two components, dissipation and feedback. The dissipation component played a large part in the restrictions of ENAQT. Due to its decoherent nature, it reduced the coherence within the system. With respect to the feedback component, it also played a part in the destruction of the system coherence for warm systems. However, its contribution is difficult to quantify and its personal effective is limited compared to the improved effective dissipation strength ($N\Gamma_F$ vs $(N+1)\Gamma_D$). However, we know it played a part due to the increase in minimum $R$ when compared to just the effective dissipation strength.

The addition of an ancillary system offered the possibility to explore different environmental regimes. When studying coherent coupling between the ancillary system and the network, evidence of non-Markovianity were found: a communal ancilla resulted in larger SEP than for the individual-ancilla counterpart. This can be understood as the provision, by the common ancilla, of an extra pathway for the excitations to transfer across. In this respect, non-Markovianity does not appear to play a crucial role in the establishment of the efficiency of the transfer, which is instead strongly dependent on the {network configuration, and especially its degree of connectivity.} Our work addresses an open problem of great technological relevance: the identification of environment-assisted effects that are enhanced by {a suitably chosen network configuration. These effects can be useful to design more efficient strategies for energy and excitation-transport in artificial nanostructures, thus emulating the behavior of biological processes such a photosynthesis.} 

%\ls{I think the next paragraph is redundant} Our study contributes to the understanding of environment-induced advantages provided to otherwise fully coherent processes, highlighting configuration-related effects that were previously overlooked. In this regard, this work contributes to the ongoing efforts for a full theoretical characterisation of processes (such as light harvesting) that can be modelled in terms of the dynamics of complex interacting quantum systems.

%\begin{itemize}  
%\item System requirements for the ENAQT effect to occur
%\item The role of the Markovian environment
%\item The effect of adding an ancillary system
%\item The difference between incoherent and coherent coupling 
%\item Non-Markovianity
%\end{itemize}

\acknowledgments

We acknowledge financial support from the Northern Ireland DfE, the EU FP7 Collaborative Project TherMiQ (grant agreement 618074), the Julian Schwinger Foundation (grant number JSF-14-7-0000), and the SFI-DfE Investigator Programme grant (grant 15/IA/2864). LS thank the EU Commission Marie Curie RISE Program (ÒENACTÓ project, grant number 643998) and the UK EPSRC (grant nr. EP/P016960/1) for funding.


\begin{thebibliography}{99}
\bibitem{Gammaitoni} L. Gammaitoni, P. H\"anggi, P. Jung, and F. Marchesoni, Rev. Mod Phys. {\bf 70}, 223 (1998). 
\bibitem{Haenggi} P. H\"anggi, Chemphyschem. {\bf 3}, 285 (2002); L. Gammaitoni, P. H\"anggi, P. Jung, and F. Marchesoni, Europ. Phys. J. B {\bf 69}, 1 (2009)
\bibitem{Plenio1} M. B. Plenio, and S. F. Huelga, Phys. rev. lett. {\bf 88}, 197901 (2002).
\bibitem{Hartmann} L. Hartmann, W. D\"ur, and H. J. Briegel, Phys. Rev. A {\bf 74}, 052304 (2006).
\bibitem{Sink1} M. Mohseni, P. Rebentrost, S. Lloyd, and A. Aspuru-Guzik, J. Chem. Phys. {\bf 129}, 174106 (2008).
\bibitem{Huelga} M. B. Plenio and S. F. Huelga, New J. Phys. {\bf 10}, 113019 (2008).
\bibitem{Chin} A. W. Chin, A. Datta, F. Caruso, S. F. Huelga, and M. B. Plenio, New J. Phys. {\bf 12}, 065002 (2010); F. Caruso, A. W. Chin, A. Datta, S. F. Huelga, and M. B. Plenio, J. Chem. Phys. {\bf 131}, 105106 (2009).
\bibitem{OlayaCastro} A. Olaya-Castro, C. F. Lee, F. Fassioli Olsen, and N. F. Johnson, Phys. rev. B {\bf 78}, 085115 (2008); F. Novelli, A. Nazir, G. H. Richards, A. Roozbeh, K. E. Wilk, P. M. G. Curmi, and J. A. Davis, arXiv:1503.00251 (2015); F. L. Semi\~ao, K. Furuya, and G. J. Milburn, New J. Phys. {\bf 12}, 083033 (2010).
\bibitem{HuelgaPlenio} S. F. Huelga, and M. B. Plenio, Contemp. Phys. {\bf 54}, 181 (2013).
\bibitem{Checinska} A. Checinska, F. A. Pollock, L. Heaney, and A. Nazir, J. Chem. Phys. {\bf 142}, 025102 (2015). 
\bibitem{Caruso} F. Caruso, A. W. Chin, A. Datta, S. F. Huelga, and M. B. Plenio, Phys. Rev. A {\bf 81}, 062346 (2010).
\bibitem{Sarovar} M. Sarovar, A. Ishizaki, G. R. Fleming, K. B. Felming, and K. B. Whaley, Nature Phys. {\bf 6}, 462 (2010).
\bibitem{Bradler} K. Bradler, M. M. Wilde, S. Vinjanampathy, and D. B. Uskov, Phys. Rev. A {\bf 82}, 062310 (2010).
\bibitem{Cifuentes} A. A. Cifuentes, and F. L. Semi\~ao, Phys. Rev. A {\bf 95}, 062302 (2017).
\bibitem{ENAQT} P. Rebentrost, M. Mohseni, I. Kassal, S. Lloyd, and A. Aspuru-Guzik, New J. Phys. {\bf 11}, 033003 (2009).
\bibitem{Lindderiv} C. A. Brasil, F. F. Fanchini, and R. de Jesus Napolitano, arXiv:1110.2122 (2012).
\bibitem{Badsink} D. Gelbwaser-Klimovsky and A. Aspuru-Guzik, arXiv:1605.04875 (2016).
\bibitem{FMO} G. S. Engel, T. R. Calhoun, E. L. Read, T.-K. Ahn, T. Mancal, Y.-C. Cheng, R. E. Blankenship, and G. R. Fleming, Nature {\bf 446}, 782 (2007).
\bibitem{FMOB} G. Panitchayangkoon, D. V. Voronine, D. Abramavicius, J. R. Caram, N. H. C. Lewis, S. Mukamel, and G. S. Engel, Proc. Natl. Acad. Sci. USA {\bf 108}, 20908 (2011).
\bibitem{QuantumBio} S. Lloyd, J. Phys.: Conf. Ser. {\bf 302}, 012037 (2011).
\bibitem{Coherence} T. Baumgratz, M. Cramer, and M. B. Plenio, Phys. Rev. Lett. {\bf 113}, 140401 (2014).
\bibitem{BPL} H.-P. Breuer, J. Piilo, and E.-M. Laine, Phys. Rev. Lett. {\bf 103}, 210401 (2009). 
\bibitem{review} H.-P. Breuer, E.-M. Laine, J. Piilo, and B. Vacchini, Rev. Mod. Phys. {\bf 88}, 021002 (2016).
\end{thebibliography}
\end{document}